\documentclass[prd,nofootinbib,preprintnumbers,twocolumn,
showpacs,floatfix,superscriptaddress]{revtex4-1}
 
\usepackage{verbatim}

\usepackage[normalem]{ulem}
\usepackage{relsize}
\usepackage{amsmath}
\usepackage{amssymb}
\usepackage{epsfig}
\usepackage{graphicx}
\usepackage{hyperref}
\usepackage{dcolumn}   
\usepackage{boldline}
\usepackage{slashed}
\usepackage{multirow}
\usepackage{color}
\usepackage[normal]{subfigure}
\usepackage{rotating}
\usepackage[margin=0.9in,a4paper]{geometry}
\usepackage[table]{xcolor}
\usepackage{enumitem}
\usepackage[utf8x]{inputenc}
\usepackage{colortbl}
\usepackage{array,multirow}
\definecolor{nicered}{rgb}{0.7,0.1,0.1}
\definecolor{niceblue}{rgb}{0.1,0.1,0.8}
\definecolor{nicegreen}{rgb}{0.1,0.5,0.1}
\definecolor{red}{rgb}{1.0, 0, 0}
\def\mX{{\mathcal{X}}}

\makeatletter

    \def\CT@@do@color{%
      \global\let\CT@do@color\relax
            \@tempdima\wd\z@
            \advance\@tempdima\@tempdimb
            \advance\@tempdima\@tempdimc
    \advance\@tempdimb\tabcolsep
    \advance\@tempdimc\tabcolsep
    \advance\@tempdima2\tabcolsep
            \kern-\@tempdimb
            \leaders\vrule
                    \hskip\@tempdima\@plus  1fill
            \kern-\@tempdimc
            \hskip-\wd\z@ \@plus -1fill }
    \makeatother

\def\eq#1{{Eq.~(\ref{#1})}}

\def\fig#1{{Fig.~\ref{#1}}}

\def\Table#1{{Table~\ref{#1}}}

\def\Sect#1{{Section~\ref{#1}}}
\def\sect#1{{section~\ref{#1}}}

\def\app#1{{Appendix~\ref{#1}}}

\def\vev#1{\left\langle #1\right\rangle}
\def\abs#1{\left| #1\right|}

\def\Tr{\mbox{Tr}\,}

\def\zN{\mathbb{Z}_\mathbb{N}}
\def\gsim{\raise0.3ex\hbox{$\;>$\kern-0.75em\raise-1.1ex\hbox{$\sim\;$}}}
\def\lsim{\raise0.3ex\hbox{$\;<$\kern-0.75em\raise-1.1ex\hbox{$\sim\;$}}}
\def\q{{\cal Q}}

\def\mb[#1]{\mathbf{#1}}

\renewcommand{\bar}{\overline}

\definecolor{LightCyan}{rgb}{0.88,1,1}
\definecolor{piggypink}{rgb}{0.99, 0.87, 0.9}
\definecolor{applegreen}{rgb}{0.55, 0.71, 0.0}
\definecolor{darkpastelgreen}{rgb}{0.01, 0.75, 0.24}
\definecolor{green-yellow}{rgb}{0.68, 1.0, 0.18}

\newcommand{\beq}{\begin{equation}}
\newcommand{\eeq}{\end{equation}}
\newcommand{\beqa}{\begin{eqnarray}}
\newcommand{\eeqa}{\end{eqnarray}}

\def\xu#1{\mX_{u_#1}}
\def\xd#1{\mX_{d_#1}}
\def\xe#1{\mX_{e_#1}}

\newcommand{\eqn}[1]{eq.~(\ref{#1})}

\begin{document}


\title{Window for preferred axion models}

\author{Luca Di Luzio}
\email{luca.di-luzio@durham.ac.uk}
\affiliation{\normalsize \it 
Institute for Particle Physics Phenomenology, Department of Physics, Durham University, DH1 3LE, United Kingdom}
\author{Federico Mescia}
\email{mescia@ub.edu}
\affiliation{\normalsize\it Dept.~de F\'{\i}sica Qu\`antica i
  Astrof\'{\i}sica, Institut de Ci\`encies del Cosmos (ICCUB), 
Universitat de Barcelona, Mart\'i Franqu\`es 1, E08028 Barcelona, Spain
}
\author{Enrico Nardi}
\email{enrico.nardi@lnf.infn.it}
\affiliation{\normalsize\it INFN, Laboratori Nazionali di Frascati, C.P.~13, 100044 Frascati, Italy}

\begin{abstract} \noindent

  We discuss phenomenological criteria for defining ``axion windows'',
  namely regions in the parameter space of the axion-photon coupling
  where realistic models live.  Currently, the boundaries of this
  region depend on somewhat arbitrary criteria, and it would be highly
  desirable to specify them in terms of precise phenomenological
  requirements. We first focus on hadronic axion models within
    post-inflationary scenarios, in which the initial abundance of the
    new vectorlike quarks $Q$ is thermal. We classify their
    representations $R_Q$ by requiring that $i)$ the $Q$ are
    sufficiently short lived to avoid issues with long-lived strongly
    interacting relics, $ii)$ the theory remains weakly coupled up to
    the Planck scale. The more general case of multiple $R_Q$ is also
    studied, and the absolute upper and lower bounds on the
    axion-photon coupling as a function of the axion mass is
    identified. Pre-inflationary scenarios in which the axion decay
    constant remains bounded as $f_a\leq 5\cdot 10^{11}\,$GeV allow
    for axion-photon couplings only about 20\% larger. Realistic
    Dine-Fischler-Srednicki-Zhitnitsky type of axion models also
    remain encompassed within the hadronic axion window.  Some
  mechanisms that can allow to enhance the axion-photon coupling to
  values sizeably above the preferred window are discussed.

\end{abstract}


\maketitle


\section{Introduction}
\label{intro}
It is well known that the standard model (SM) of particle physics does
not explain some well established experimental facts like dark matter
(DM), neutrino masses, and the cosmological baryon asymmetry, and it
also contains fundamental parameters with highly unnatural values,
like the coefficient $\mu^2\sim O((100\ \text{GeV})^2)$ of the quadratic
term in the Higgs potential, the Yukawa couplings of the first family
fermions $h_{e,u,d}\sim 10^{-6} - 10^{-5}$ and the strong CP violating
angle $|\theta| < 10^{-10}$.  This last quantity is somewhat special:
its value is stable with respect to higher order corrections
\cite{Ellis:1978hq} (unlike $\mu^2$) and (unlike
$h_{e,u,d} $~\cite{Hall:2014dfa}) it evades explanations based on
environmental selection~\cite{Ubaldi:2008nf}.  Thus, seeking
explanations for the smallness of $\theta$ independently of other
``small values'' problems is theoretically motivated.  While most of
the  problems of the SM can be addressed with a large variety of mechanisms,
basically only three types of solutions to the strong CP problem have
been put forth so far.  The simplest possibility, a massless up-quark,
is now ruled out ($m_u\neq 0$ by 20 standard
deviations~\cite{Aoki:2013ldr,Manohar}).  The so-called Nelson-Barr
(NB) type of models~\cite{Nelson:1983zb,Barr:1984qx} either require a
high degree of fine tuning, often comparable to setting
$|\theta| \lsim 10^{-10}$ by hand, or additional and rather elaborated
theoretical structures to keep $\theta$ sufficiently small at all
orders~\cite{Vecchi:2014hpa,Dine:2015jga}. The Peccei-Quinn (PQ)
solution~\cite{Peccei:1977ur,Peccei:1977hh} arguably stands on better
theoretical grounds, and from the experimental point of view it also
has the advantage of predicting an unmistakable signature: the
existence of a new light scalar particle, universally known as the
axion~\cite{Weinberg:1977ma,Wilczek:1977pj}.  Therefore, the issue if
the PQ solution is the correct one, could be set experimentally by
detecting the axion. In contrast, no similar unambiguous
signature exists for NB models.

A crucial challenge for axion models is to explain through which
mechanism the global $U(1)_{PQ}$ symmetry, on which the solution
relies (and that presumably arises as an accident), remains protected
from explicit breaking to the required level of
accuracy~\cite{Kamionkowski:1992mf,Holman:1992us,Barr:1992qq}, and it
seems fair to state that only constructions that embed such an
explanation can be considered theoretically satisfactory.  A wide
variety of proposals to generate a high quality $U(1)_{PQ}$ have been
put forth based, for example, on discrete gauge
symmetries~\cite{Dine:1992vx,Dias:2002gg,Dias:2014osa,Ringwald:2015dsf},
supersymmetry~\cite{Holman:1992us,Harigaya:2013vja,Festuccia:2015kra},
compositeness~\cite{Randall:1992ut,Dobrescu:1996jp,Redi:2016esr,Dvali:2016eay}, 
flavour symmetries \cite{Cheung:2010hk} or new continuous gauge symmetries~\cite{Fukuda:2017ylt,DiLuzio:2017tjx}.
Regardless of the details of the different theoretical constructions,
many properties of the axion remain remarkably independent from
specific model realizations.  It is then very important, in order to
focus axion searches, to identify as well as possible the region in
parameter space where realistic axion models live. The vast majority
of axion search techniques are sensitive to the axion-photon coupling
$g_{a\gamma\gamma}$ which is inversely proportional to the axion decay
constant $f_a$.  Since the axion mass $m_a$ has the same dependence,
the experimental exclusion limits, as well as the theoretical predictions
for specific models, can be conveniently presented in the
$m_a$-$g_{a\gamma\gamma}$ plane (see Fig.~\ref{fig:Excl}).  The
commonly adopted 
``axion band'' corresponds roughly to
\beq
g_{a\gamma\gamma} \sim  \frac{\alpha}{2\pi} \frac{m_a}{f_\pi m_\pi} \sim
\frac{10^{-10}}{\text{GeV}}\,\left(\frac{m_a}{{\rm eV}}\right) \,,
\eeq
with a somewhat arbitrary width chosen to include representative
models as e.g.~those of
Refs.~\cite{Kaplan:1985dv,Cheng:1995fd,Kim:1998va}. Recently, in
Ref.~\cite{DiLuzio:2016sbl} we have put forth a definition of a {\it
  phenomenologically preferred} axion window as the region
encompassing {\it hadronic} axion models which $i)$ do not contain
cosmologically dangerous strongly interacting relics; $ii)$ do not
induce Landau poles (LP) below a scale $\Lambda_{LP}$ of the order of
the Planck scale. In this paper we will first present a more detailed
analysis of the phenomenological constraints on hadronic axion models
(to which we will often refer also as Kim-Shifman-Vainshtein-Zakharov
(KSVZ)~\cite{Kim:1979if,Shifman:1979if} type of axion models) on which
the study of Ref.~\cite{DiLuzio:2016sbl} was based.
Since the first condition $i)$ is relevant only when the heavy
  quarks $Q$ have an initial thermal abundance, the validity of the
  analysis in Ref.~\cite{DiLuzio:2016sbl} is restricted to the case
  when $T_{\rm reheating} \gtrsim m_Q$. The $Q$ acquire their mass via
  a Yukawa coupling with the complex axion field so that, for Yukawa
  couplings not exceeding unity, this translates into
  $T_{\rm reheating} \gtrsim f_a$ (where $f_a$ is the axion decay
  constant) a condition that can be only realized when the PQ symmetry
  is broken after inflation, and will be referred as {\it
    post-inflationary scenario}.  However, astrophysical
  considerations imply a lower bound $f_a \gtrsim 10^9\,$GeV, while
  the only firm limit on the scale of inflation is provided by big
  bang nucleosynthesis (BBN) to merely lie above a few MeV. Since this
  leaves ample space for axion models to be realized in
  pre-inflationary scenarios, in which the initial $Q$ abundance is
  completely negligible, it would be interesting to generalize the
  analysis of~\cite{DiLuzio:2016sbl} by dropping condition $i)$.  Such
  a generalization will be carried out in \sect{preinflationary},
  subject to the only condition that $f_a \leq 5\times 10^{11}\,$GeV,
  which restricts the class of models to those which do not require
  any ad hoc tuning (or anthropic selection arguments) to justify
  particularly small initial values of $\theta$.  As we will show, in
  pre-inflationary scenarios the LP condition {\it ii)} alone is
  sufficiently strong that the limits found in~\cite{DiLuzio:2016sbl}
  get relaxed at most by $\approx 20\%$. 
In \sect{sec:DFSZ} we extend the analysis to include also the
Dine-Fischler-Srednicki-Zhitnitsky (DFSZ)
axion~\cite{Zhitnitsky:1980tq,Dine:1981rt}, that was not considered
in~\cite{DiLuzio:2016sbl}, together with several of
its variants, to which we will collectively refer as DFSZ-type of
models.  We will argue that the same window that encompasses preferred
hadronic axion models, also includes the majority of realistic DFSZ
scenarios.

The layout of the paper is the following: in \sect{hadronicaxions} we
introduce hadronic axion models with some focus on the issue of the
stability of the new heavy quarks $Q$. \Sect{sec:cosmo} is devoted to
the cosmological consequences of stable or long-lived $Q$'s: we
estimate the present abundances of strongly interacting relics arguing
that absolutely stable $Q$'s are likely excluded, and we review the
constraints on the lifetimes of metastable $Q$'s.  In
\sect{axionwindow} we put forth a definition of preferred hadronic
(KSVZ) axion models on the basis of our two well-defined
discriminating criteria.  In~\sect{axioncoupling} we identify the
window in parameter space where preferred axion models live, and we
discuss the corresponding maximum and minimum values allowed for the
axion-photon coupling.  In~\sect{preinflationary} we address KSVZ
  models in pre-inflationary scenarios showing that the previous
  results get only mildly relaxed as long as the requirement
  $f_a \leq 5\times 10^{11}\,$GeV on the axion decay constant is
  maintained.  In~\sect{sec:DFSZ} we address DFSZ-type of axion
models showing that the same window also includes realistic models of
this type. Finally, in \sect{concl} we review the main results and
draw the conclusions.

\section{Hadronic axions}
\label{hadronicaxions}

The basic ingredient of any axion model is a global 
$U(1)_{PQ}$ symmetry. The associated Noether current must have 
a color anomaly and, although not required for solving the strong CP
problem,  in general it also has an electromagnetic anomaly:
\begin{equation}
\label{eq:NE}
\partial^\mu J_\mu^{PQ} = 
\frac{N \alpha_s}{4\pi} G\cdot\tilde G + 
\frac{E \alpha}{4\pi} F\cdot\tilde F\,,
\end{equation}
where $G,\,F$ are the color and electromagnetic field strength
tensors, $\tilde G,\,\tilde F$ their 
duals (e.g. $F\cdot\tilde F \equiv \frac{1}{2} \epsilon_{\mu\nu\rho\sigma} F^{\mu\nu} F^{\rho\sigma}$, etc.), 
and $N$ and $E$
are the color and electromagnetic anomaly coefficients.  
In a generic
axion model of KSVZ type~\cite{Kim:1979if,Shifman:1979if} the anomaly
is induced by pairs of 
heavy fermions $Q_L,\,Q_R$ which must
transform non-trivially under $SU(3)_C$ and chirally under $U(1)_{PQ}$.
Their mass arises from  a Yukawa interaction with a SM singlet
scalar field $\Phi$ which develops a 
PQ breaking vacuum expectation value. Therefore 
their PQ charges $\mX_{L,R}\equiv\mX(Q_{L,R})$, normalized to
$\mX(\Phi)=1$, must satisfy
\beq
\label{eq:chiLR}
|\mX_L-\mX_R|=1\,.  
\eeq
In KSVZ models the SM fermions do not contribute to the color or
  electromagnetic anomalies so that their PQ charges can be set to
  zero.  We denote the (vectorlike) representations of the SM gauge
group $G_{SM}\!\!=\!SU(3)_C\!\times\! SU(2)_I\!\times\! U(1)_Y$ to
which the $Q$ are assigned as
$R_Q\!\!=\! (\mathcal{C}_Q,\mathcal{I}_Q,\mathcal{Y}_Q)$, so that
the anomaly coefficients read:
\beqa 
\label{N}
N &=& \sum_Q
\left(\mX_L-\mX_R\right)\,T(\mathcal{C}_Q) \,, \\  
\label{E}
E &=& \sum_Q \left(\mX_L-\mX_R\right)\,\mathcal{Q}^2_Q \,.   
\eeqa 
Here $\sum_Q$ denotes the sum over all irreducible
$SU(3)_C \times U(1)_{\rm em}$ representations (we allow for the
simultaneous presence of more $R_Q$). The color index is defined by
$\Tr T_Q^a T_Q^b = T(\mathcal{C}_Q) \delta^{ab}$ with $T_Q$ the
generators in $\mathcal{C}_Q$ (in particular, $T(3)=1/2$, $T(6)=5/2$,
$T(8)=3$, $T(15)=10$) and $\mathcal{Q}_Q$ denotes the $U(1)_{\rm{em}}$
charge.  Different choices for $R_Q$ imply different phenomenological
consequences, and we will use this fact to identify phenomenologically
preferred models.  Let us parametrize the scalar field $\Phi$ as
\begin{equation}
\label{eq:Phi}
\Phi(x) = \frac{1}{\sqrt{2}} \left[\rho(x) + V_a\right] e^{i
  a(x)/V_a}\,. 
\end{equation} 
$\rho(x)$ acquires a mass $m_\rho \sim V_a$ while $a(x)$ is the
axion field which would remain massless in the absence of explicit
$U(1)_{PQ}$ breaking. In invisible axion models, in order to sufficiently suppress 
the axion couplings that  scale as $1/f_a\equiv 2 N/V_a$,  
it is assumed that 
$V_a \gg (\sqrt{2} G_F)^{-1/2} = 247\,$GeV. More
quantitatively, astrophysical constraints hint to a lower limit
$f_a  \gsim 4\cdot 10^8\,$GeV~\cite{Kuster:2008zz,Marsh:2015xka}. 

The renormalizable Lagrangian for a generic hadronic axion 
model can be written as:
\begin{equation}
\label{LKSVZ}
\mathcal{L}_a = 
\mathcal{L}_{\rm SM}+
\mathcal{L}_{\rm PQ} 
 - V_{H \Phi} +
\mathcal{L}_{Qq} 
\,,
\end{equation}
where $\mathcal{L}_{\rm SM}$  is the SM Lagrangian, 
\begin{equation}
\label{LKSVZPQ}
\mathcal{L}_{\rm PQ} = |\partial_\mu \Phi|^2 + 
\overline{Q} i \slashed{D} Q 
- (y_Q \, \overline{Q}_L Q_R \Phi + \text{H.c.}) \, ,
\end{equation}
with $Q=Q_L+Q_R$ and, from the last term,  
$m_Q =y_Q V_a/\sqrt{2}$. 
$V_{H \Phi}$ contains 
the new scalar couplings:
\begin{equation}
\label{LKSVZV}
V_{H \Phi} 
= 
- \mu^2_\Phi  |\Phi|^2   + \lambda_\Phi |\Phi|^4 +  \lambda_{H\Phi}
|H|^2 |\Phi|^2 \,.
\end{equation} 
Finally, $\mathcal{L}_{Qq}$ contains possible renormalizable terms 
coupling  $Q_{L,R}$ to the
SM quarks $q=q_L,d_R,u_R$, which can allow $Q$ 
decays~\cite{Ringwald:2015dsf}. 
Note however, that SM gauge invariance allows $\mathcal{L}_{Qq} \neq 0$ only for
few specific $R_Q$ and, for example, the original KSVZ assignment
$R_Q=(3,1,0)$~\cite{Kim:1979if,Shifman:1979if} implies
$\mathcal{L}_{Qq} = 0$ (and it would in fact forbid $Q$-decays to all
orders).

\subsection{$Q$ stability and PQ quality}  
\label{accidentalsym}

The issue whether the $Q$'s are exactly stable, metastable, or decay
with safely short lifetimes, is of central importance for KSVZ models
in post-inflationary scenarios, and we will now discuss it in more
detail.  The gauge invariant kinetic term in $\mathcal{L}_{\rm PQ}$
possesses a
$U(1)^3 \equiv U(1)_{Q_L} \times U(1)_{Q_R} \times U(1)_{\Phi}$
symmetry corresponding to independent rephasing of the $Q_{L,R}$ and
$\Phi$ fields.  The PQ Yukawa term ($y_Q\neq 0$) breaks
$U(1)^3 \to U(1)_{PQ}\times U(1)_{Q}$ where, in analogy to ordinary
baryon number $U(1)_B$ for the SM quarks, $U(1)_Q$ is the $Q$-baryon
number of the new quarks~\cite{Kim:1979if} under which
$Q_{L,R} \rightarrow e^{i \beta} Q_{L,R}$ and $\Phi \rightarrow \Phi$.
Moreover, $U(1)_Q$ being vectorlike is not broken by anomalies.  If it
were an exact symmetry, $U(1)_Q$ would ensure absolute $Q$ stability,
a possibility which is preferable to avoid.  In the few cases in which
$\mathcal{L}_{Qq}\neq 0$ is allowed by $G_{SM}$ gauge invariance,
$U(1)_Q\times U(1)_B$ is further broken to $U(1)_{B'}$, that is a
generalized baryon number extended to the $Q$'s, which can then decay
into SM quarks with unsuppressed rates.  However, whether
$\mathcal{L}_{Qq}$ is allowed at the renormalizable level, does not
depend solely on $R_Q$, but also on the specific PQ charges. For
example, independently of $R_Q$, if $U(1)_{PQ}$ were an exact
symmetry, the common assignment $\mX_L=-\mX_R=\frac{1}{2}$ would
forbid PQ invariant decay operators of any dimension.  More
realistically, both $U(1)_{PQ}$ and $U(1)_Q$ are expected to be broken
at least by Planck-scale effects, inducing PQ violating contributions
to the axion potential $V^{d>4}_{\Phi}$ as well as an effective
Lagrangian $\mathcal{L}^{d>4}_{Qq}$. In particular, in order to
preserve $\abs{\theta}<10^{-10}$, operators in $V^{d>4}_{\Phi}$ must
be of dimension
$d\geq 11$~\cite{Kamionkowski:1992mf,Holman:1992us,Barr:1992qq}, and
if $\mathcal{L}^{d>4}_{Qq}$ had to respect $U(1)_{Q}$ to a similar
level of accuracy, then the $Q$'s would behave as effectively
stable. However, a scenario in which the global $U(1)_Q$ symmetry
arises as an accident because of specific assignments for the charges
of another global symmetry $U(1)_{PQ}$ seems theoretically untenable.
It would be instead desirable to enforce on the basis of first
principles a situation in which $(i)$ $U(1)_{PQ}$ arises accidentally
and is of the required {\it high quality}, $(ii)$ $U(1)_Q$ is either
broken at the renormalizable level, or it can be of a sufficient {\it
  bad quality} to allow for sufficiently fast $Q$ decays.  Here we
will not commit ourselves to any specific mechanism to realize such a
scenario, and we will simply adopt a technical solution to this issue:
a discrete (gauge) symmetry $\zN$ under which
$\Phi \rightarrow \omega\, \Phi$ (with
$\omega \equiv e^{i 2 \pi/\mathbb{N}}$) which can automatically ensure
that the minimum dimension of the PQ breaking operators in
$V^{d>4}_{\Phi}$ is $\mathbb{N}$, so that the first condition is
satisfied if $\mathbb{N}\geq 11$. In Ref.~\cite{DiLuzio:2016sbl} it
was shown that at the same time suitable transformations for $Q_{L,R}$
under $\zN$ can be found that allow the $Q$'s to decay via operators
of much lower dimension $d\leq 5$. Although, admittedly, such a
solution seems just as an ad hoc construction, it suffices to ensure
that it is consistent to assume that a high quality $U(1)_{PQ}$ can
live together with a $U(1)_Q$ of sufficiently bad quality.

\section{Heavy quarks cosmology}
\label{sec:cosmo}

We start by assuming a post-inflationary scenario ($U(1)_{PQ}$ broken
after inflation).  In this case, requiring that the axion energy
density from vacuum realignment does not exceed $\Omega_{DM}$ implies
$f_a \lesssim 5\cdot 10^{11}$
GeV~\cite{Bonati:2015vqz,Petreczky:2016vrs,Borsanyi:2016ksw}.  We
further assume $m_Q < T_{\rm reheating}$ so that via gauge
interactions the $Q$'s will attain a thermal distribution, providing
the initial conditions for their cosmological history, which will then
depend only on their mass $m_Q$ and representation $R_Q$.

For some $R_Q$ the heavy quark can only hadronize into fractionally
charged hadrons, and in this case, as detailed in
\app{app:integer-charged}, decays into SM particles are
forbidden. These $Q$-hadrons must then exist today as stable relics.
Searches for fractionally charged particles limit their abundance with
respect to ordinary nucleons to
$n_Q/n_b \lsim 10^{-20}$~\cite{Perl:2009zz}. This is orders of
magnitude below any reasonable estimate of their relic abundance and
of their concentrations in bulk matter.  This restricts the possible
$R_Q$'s to the much smaller subset which allows for integrally charged
or neutral color singlet $Q$-hadrons, in which case the limits on
cosmologically stable heavy relics are less tight.  However, for each
$R_Q$ belonging to this subset it is always possible to construct
gauge invariant operators that can allow the $Q$ to decay into SM
particles (see \app{app:integer-charged}).  Let us start by discussing
the case of lifetimes $\tau_Q$ shorter than the age of the Universe,
so that no heavy relics are left around during the present era.
Cosmological observations severely constrains the allowed values for
$\tau_Q$.  For $\tau_Q \sim (10^{-2}-10^{12})\,$s the decays of
superheavy quarks with $m_Q\gg 1\,$TeV would affect
BBN
\cite{Kawasaki:2004qu,Jedamzik:2006xz,Jedamzik:2007qk,Kawasaki:2017bqm}.
Early energy release from heavy particles decays with lifetimes
$\sim (10^{6}-10^{12})$ s is strongly constrained also by limits on
CMB spectral
distortions~\cite{Hu:1993gc,Chluba:2011hw,Chluba:2013wsa}, while $Q$'s
decaying around the recombination era ($t_{\rm rec} \sim 10^{13}\,$s)
are tightly constrained by measurements of CMB anisotropies.  Decays
after recombination would give rise to free-streaming photons visible
in the diffuse gamma ray background \cite{Kribs:1996ac}, and tight
constraints from Fermi LAT \cite{Ackermann:2012qk} allow to exclude
$\tau_Q\sim (10^{13}-10^{26})\,$s. Note that these last constraints
are also able to exclude lifetimes that are several order of magnitude
larger than the age of the Universe, $t_U\sim 4\cdot 10^{17}\,$s.

\subsection{Abundance of strongly interacting relics}

Cosmologically stable $Q$'s are severely constrained by the
requirement that their present energy density does not exceed that of
the DM $\Omega_Q \leq \Omega_{DM} \sim 0.12\, h^{-2}$.  Obtaining
reliable estimates of $\Omega_Q$ is a non-trivial task though. Some
controversy in the results exists, mainly related to possible large
enhancements of the annihilation rate with respect to free $Q$'s
annihilation, which can occur after the $Q$'s get confined into
hadrons.  We now review the state of the art, and we formulate a
motivated guess about the most reasonable range of values for
$\Omega_Q$.

\begin{figure}[t!]
\centering
\includegraphics[angle=0,width=7.9cm]{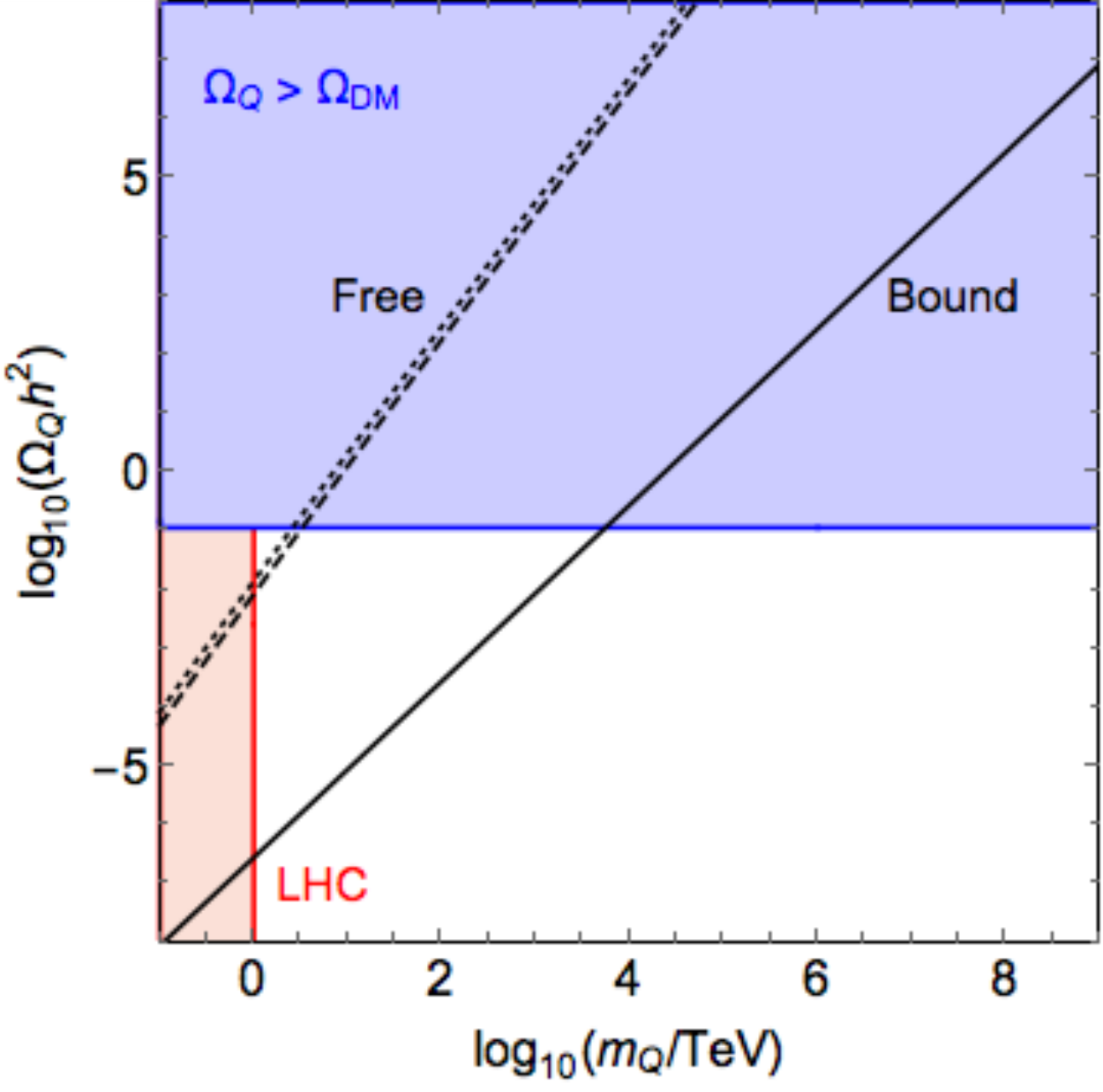}
\caption{\label{fig:overclosure} Contribution to the cosmological
  energy density versus $m_Q$.  The broken lines correspond
  to free $Q$ annihilation for color triplets (dotted) and octets
  (dashed). The solid line to annihilation below $T_C$
  via bound state formation.  
The horizontal and vertical lines $\Omega_Q= \Omega_{DM}$ and 
$m_Q=1\,$TeV limit the allowed region.}
\end{figure}

At temperatures above the QCD phase transition $T_{C} \sim 180\,$MeV
the $Q$'s annihilate as free quarks.  One generally assumes a
symmetric scenario $n_Q=n_{\bar Q}$ since any asymmetry would
eventually quench $Q\bar Q$ annihilation resulting in stronger bounds.
Perturbative computations in this regime are reliable, and give:
\begin{equation}
  \label{eq:pertannh}
\langle \sigma v\rangle_{Q\bar Q} = 
  \frac{\pi\alpha_s^2}{16 m_Q^2}\left(c_f n_f + c_g\right) \,, 
\end{equation}
where $n_f$ is the number of quark flavors into which $Q$ can
annihilate, and $(c_f,c_g)= (\frac{2}{9},\frac{220}{27})$ for triplets \cite{Olive:2016xmw},
$(\frac{3}{2},\frac{27}{4})$ for octets \cite{Baer:1998pg}, etc.
Freezeout of free $Q$ annihilation occurs around $T_{fo} \sim m_Q/25$
and, for $m_Q>$ few TeV, at  $T_{fo}$ there are $g_* = 106.75$ effective degrees of
freedom in thermal equilibrium. Together with~\eq{eq:pertannh} this
gives:
\begin{align}
\label{relicpertQCD}
\left( \Omega_{Q} h^2 \right)^{\rm Free} 
&=2.0 
\left( \frac{10^{-10} \ \rm{GeV}^{-2}}{\vev{\sigma v}_{Q\bar Q}} \right) \nonumber \\
&\approx 7.8 \cdot 10^{-3}  \left(\frac{m_Q}{{\rm TeV}}\right)^2\,, 
\end{align}
where the second equality holds for color triplets and for reference
values of the relevant parameters.  The upper lines in
\fig{fig:overclosure} give $\left( \Omega_{Q} h^2 \right)^{\rm Free}$
as a function of $m_Q$ for $SU(3)_C$ triplets (dotted) and octets
(dashed), including the running of $\alpha_s (\mu = m_Q)$ computed at
two-loops.  We see that only in a narrow interval at low $m_Q$ the
limit $\Omega_Q\leq \Omega_{DM} \sim 0.12 \, h^{-2}$ is respected, and
an improvement of the lower limit on $m_Q$ by a factor of a few
would exclude also this region.

For $T < T_{C}$ the $Q$'s get confined in color singlet hadrons, that
can be pictured as a heavy parton surrounded by a QCD cloud (``brown
muck'') of light degrees of freedom.  As the temperature decreases
below $T_C$, the presence of a baryon asymmetry for the SM quarks
implies that the brown muck is preferentially constituted by light
quarks $q$ (and eventually gluons $g$) rather than by antiquarks
$\bar q$. For example, for a color triplet, the heavy meson $Q\bar q$
will readily scatter with ordinary nucleons which are relatively much
more abundant, giving rise to a heavy baryon:
$Q\bar q + qqq \to Qqq + \bar q q$.  For $Q$'s belonging to different
$SU(3)_C$ representations different types of color singlet baryons and
mesons, including exotic and hybrid states, will eventually form, for example:
$ Qq\    (Q\in \mathbf{\bar 3}),  \ 
Qqq\ (Q\in \mathbf{3},\, \mathbf{\overline{6}}),\ 
Qqqq\ (Q \in \mathbf{8},\, \mathbf{\overline{10}}), \
 Qgg\ (Q \in \mathbf{10},\, \mathbf{\overline{10}}), $ etc. 
 However, the most important feature concurring to determine possible
 large enhancements of the annihilation cross section after
 confinement is largely independent of many fine details, and is
 essentially related to the fact that the QCD cloud of light quarks
 and gluons surrounding the heavy partons results in composite states
 of typical hadronic size $R_h\sim\,$fm.  Then, because of finite size
 effects of the hosting composite state, $Q$ annihilation can restart
 below $T_C$, and the relic abundance of $Q$-hadrons can get further
 depleted until a new freezeout temperature 
 is reached.
Clearly, if this picture is
 correct, annihilation in the perturbative regime will be to a large
 extent irrelevant, and the relic density of $Q$-hadrons will be
 essentially fixed by non perturbative processes occurring at $T<T_C$.
 Presently, agreement on quantitative estimates of the annihilation
 rate for massive colored particles confined into hadrons has not
 been reached, and published estimates for the relic density span over
 several orders of magnitude.  This issue is of central
 importance for the present study so that,  after reviewing  
the relevant literature, we will  attempt to pin down some general
 conclusion.

 In Ref.~\cite{Dover:1979sn} the relic density of confined heavy
 stable color sextet quarks was estimated by assuming an annihilation
 cross section of typical hadronic size
 $\sigma_{\rm ann} \sim (m_\pi^2 v)^{-1} \sim 30 v^{-1}\,$mb. This
 resulted in $n_Q/n_b \sim 10^{-11}$, where $n_b $ is the present
 abundance of baryons.  In~\cite{Nardi:1990ku} it was remarked that
 the quoted value of $\sigma_{\rm ann}$ was likely overestimating the
 annihilation cross section by a few orders of magnitude. This is
 because it corresponds to a typical inclusive hadronic scattering
 cross section, and it was argued that the relevant exclusive
 annihilation channel (not containing the heavy $Q$ quarks in the
 final state) would not exceed a fraction $f<1$ of the geometrical
 cross section $\sigma_r\sim \pi R_h^2$. In this case, even assuming
 $f\sim 1$, a much larger abundance
 $n_Q/n_b \sim 10^{-9} (m_Q/10\ {\rm TeV})^{-1}$ would result.  The
 relic density of heavy stable gluinos (that is $Q$'s in the adjoint
 of $SU(3)_C$) was studied in \cite{Baer:1998pg} considering several
 different possibilities for $\sigma_{\rm ann}$ ranging from
 perturbative, perturbative dressed with Sommerfeld enhancements, and
 various non-perturbative possibilities.  For the reference value
 $m_Q=10\,$TeV they found results ranging from overclosure
 $\Omega_Q h^2 \sim 1$ down to $\Omega_Q h^2 \sim 5\cdot 10^{-10}$
 which corresponds to
 $n_Q/n_b \gsim 10^{-12} (m_Q/10\,{\rm TeV})^{-1}$.

 Cosmologically long-lived gluinos were reconsidered
 in~\cite{Arvanitaki:2005fa}. The authors correctly identify the
 relevant cross section as the one characterizing, after the QCD phase
 transition, the annihilation of $Q$-hadrons. Similarly to
 Ref.~\cite{Nardi:1990ku} they argue that the expected cross section
 is not of hadronic size, but it should rather be characterized by the
 size of the heavy parton localized inside the hadron core,
 i.e. $\sigma \propto 1/m^2_Q$, a conclusion apparently supported by
 the partial wave unitarity limit for the inelastic cross
 section~\cite{Griest:1989wd}
 $\sigma^{inel}_J \leq \pi (2 J+1)/(m_Q^2 v^2)$.  However, the very
 large mass $m_Q \gg\,$TeV relative to the typical energy
 $E_Q\lsim T_C$ implies that the corresponding momentum is large, and
 many partial waves are involved in the collision.  If all partial
 waves up to $J_{max} \sim m_Q v R_h$ contribute, the same geometrical
 behavior $\sigma^{inel} = \sum \sigma^{inel}_J \sim \pi R_h^2$
 considered in~\cite{Nardi:1990ku} would be recovered.  Clearly,
 $\sigma^{inel}$ is not by itself the $Q\bar Q$ annihilation cross
 section, but includes all inelastic processes as e.g.~the formation
 of bound states out of the collision of two $Q$-hadrons, apparently
 supporting the conclusion that $f<1$.  However, if the bound states
 can efficiently radiate off large amounts of angular momentum, the
 $Q$ and $\bar Q$ wavefunctions will eventually be brought to overlap
 so that also bound state formation would contribute to annihilation.
 Collapse to states of low angular momentum must however occur in a
 relatively short time, so that annihilations will occur well before
 the BBN era. Ref.~\cite{Arvanitaki:2005fa} estimated the
 rate for angular momentum radiation via $\pi$ emission, and found it
 to be very small, concluding that bounded $Q$'s would remain
 incapable of annihilating on a sufficiently short time scale.  The
 most conservative limits quoted in that paper are then obtained under
 the assumption that the annihilation cross section saturates the $s$
 and $p$ wave unitarity limits, which for $m_Q = 10\,$TeV yields
 $n_Q/n_b \gsim 10^{-4}$, close to saturation of the cosmological
 limit $\Omega_Q < \Omega_{DM}$.

Annihilation via bound state formation was reconsidered
in~\cite{Kang:2006yd} and, as regards the radiation time to collapse
down to low angular momentum states, opposite conclusions with respect
to~\cite{Arvanitaki:2005fa} were reached: they find that for charged
$Q$-hadrons, photon radiation can collapse the bound state with a time
scale $\tau_{\rm rad} \lsim 1$ s for all masses
$m_Q\lsim 10^{11}\,$GeV (for neutral partons and neutral host
$Q$-hadrons however, this reduces to $m_Q\lsim 2.5\,$TeV).  Their
conclusion is that for charged states the relic density of $Q$'s gets
sizeably depleted by the second stage of annihilation after
hadronization. For the $Q$ contribution to the present energy density
the results of~\cite{Kang:2006yd} imply:
\begin{align}
\label{relicKLNQCD}
\nonumber
\left( \Omega_{Q} h^2 \right)^{\rm Bound} 
&\approx 
2.8\cdot 10^{-7} 
\left( \frac{R_h}{\text{GeV}^{-1}} \right)^{-2}  \qquad \\
&\times \left( \frac{T_C}{180 \ \text{MeV}} \right)^{-3/2} \left( \frac{m_Q}{\text{TeV}} \right)^{3/2} 
\!\! . 
\end{align}

The mechanism of annihilation via bound state formation was put under
closer scrutiny in~\cite{Jacoby:2007nw}, where previously neglected
effects of the large number of thermal bath pions ($n_\pi \gg n_Q$) on
the bound states were considered. The two relevant effects are
breakup of the bound states due to collisions with $\pi$'s with energy
larger than the typical binding energy $E_B \sim$ few $100\,$MeV, and
de-excitation processes through which the colliding $\pi$ carries away
two units of angular momentum.
 These two processes work in the opposite directions of delaying and
 speeding up $Q\bar Q$ annihilation, and it was estimated that
 eventually they would roughly equilibrate each other, yielding a
 result not far from the estimates in~\cite{Kang:2006yd}.

 A more quantitative study of this mechanism was carried out in
 Ref.~\cite{Kusakabe:2011hk}. The conclusion was that \eq{relicKLNQCD}
 represents a conservative lower limit on $\Omega_Q$, but that much
 larger values are also possible. In
 fact~\cite{Kang:2006yd,Jacoby:2007nw} did not consider the possible
 formation of $(QQ...)$ bound states which, opposite to $Q\bar Q$,
 would hinder annihilation rather than catalyze it.  This possibility
 was discussed in Ref.~\cite{Kusakabe:2011hk} but was not included in
 their quantitative analysis.  However, doubly and triply heavy
 baryons, like $\Omega_{ccq}$, $\Omega_{bbb}$
 (see~\cite{Shah:2016vmd,Vijande:2015faa} for compilations of recent
 results) are a firm prediction of the quark model,
also supported by the recent discovery of the doubly heavy hadron $\Xi^+_{cc}$
by the LHCb collaboration \cite{Aaij:2017ueg}.
 Clearly, the size of bound states solely
 composed by $Q$'s or by $\bar Q$'s would be much smaller than
 hadronic, quenching all enhancements of the annihilation, and if a
 relevant fraction of $Q$'s ends up in multi-$Q$ bound states, the
 final relic density would be better approximated by the free quark
 result~\eq{relicpertQCD} rather than by~\eq{relicKLNQCD}.

 By evaluating \eq{relicKLNQCD} for reference values of the relevant
 parameters, we obtain the continuous line in \fig{fig:overclosure}
 which, according to the discussion above, should be understood as a
 rather conservative lower limit on $\Omega_Qh^2$.  However, even
 assuming that the relic abundance approaches this lower
 limit, the relative concentration of $Q$-hadrons
 $n_Q/n_b \sim 10^{-8} \left(m_Q/{\rm TeV}\right)^{1/2}$ would still
 be quite large. If the $Q$'s accumulate with similar concentrations
 within the galactic disk, existing limits from searches of
 anomalously heavy isotopes in terrestrial, lunar, and meteoritic
 materials~\cite{Perl:2001xi} would be able to exclude their existence
 for most of the range of masses allowed by the $\Omega_Q<\Omega_{DM}$
 constraint.  Many other arguments have been put forth disfavoring
 the possibility of heavy stable $Q$'s: their capture in neutron stars
 would form black holes on a time scale of a few
 years~\cite{Gould:1989gw} and, more generically, they could endanger
 stellar stability~\cite{Hertzberg:2016jie}, while their annihilation
 in the Earth interior would result in an anomalously large heat flow
 \cite{Mack:2007xj}.  In conclusion, unless an extremely efficient
 mechanism exists that keeps $Q$-matter completely separated from
 ordinary matter, the possibility of stable $Q$-hadrons 
that were once in thermal equilibrium is ruled out.

\subsection{$Q$ Lifetimes} 
\label{Qlifetimes}
We have seen in the previous sections that cosmologically stable heavy
$Q$'s with $m_Q < T_{\rm{reheating}}$ are strongly disfavored,
and that in case they are unstable, only lifetimes
$\tau_Q\lsim 10^{-2}\,$s are safe with respect to cosmological issues.
For the post-inflationary case, we will then consider as
phenomenologically preferred only scenarios in which this condition
can be satisfied.  The order of magnitude of $\tau_Q$ crucially
depends on the dimension $d$ of the operators responsible for the
decays.  Below we derive quantitative estimates for $\tau_Q$ as a
function of $m_Q$ and $d$, and we argue that only for $d=4$ and $d=5$
the constraint $\tau_Q\lsim 10^{-2}\,$s can be satisfied in a natural
way.

\begin{figure}[t]
\centering
\includegraphics[angle=0,width=7.9cm]{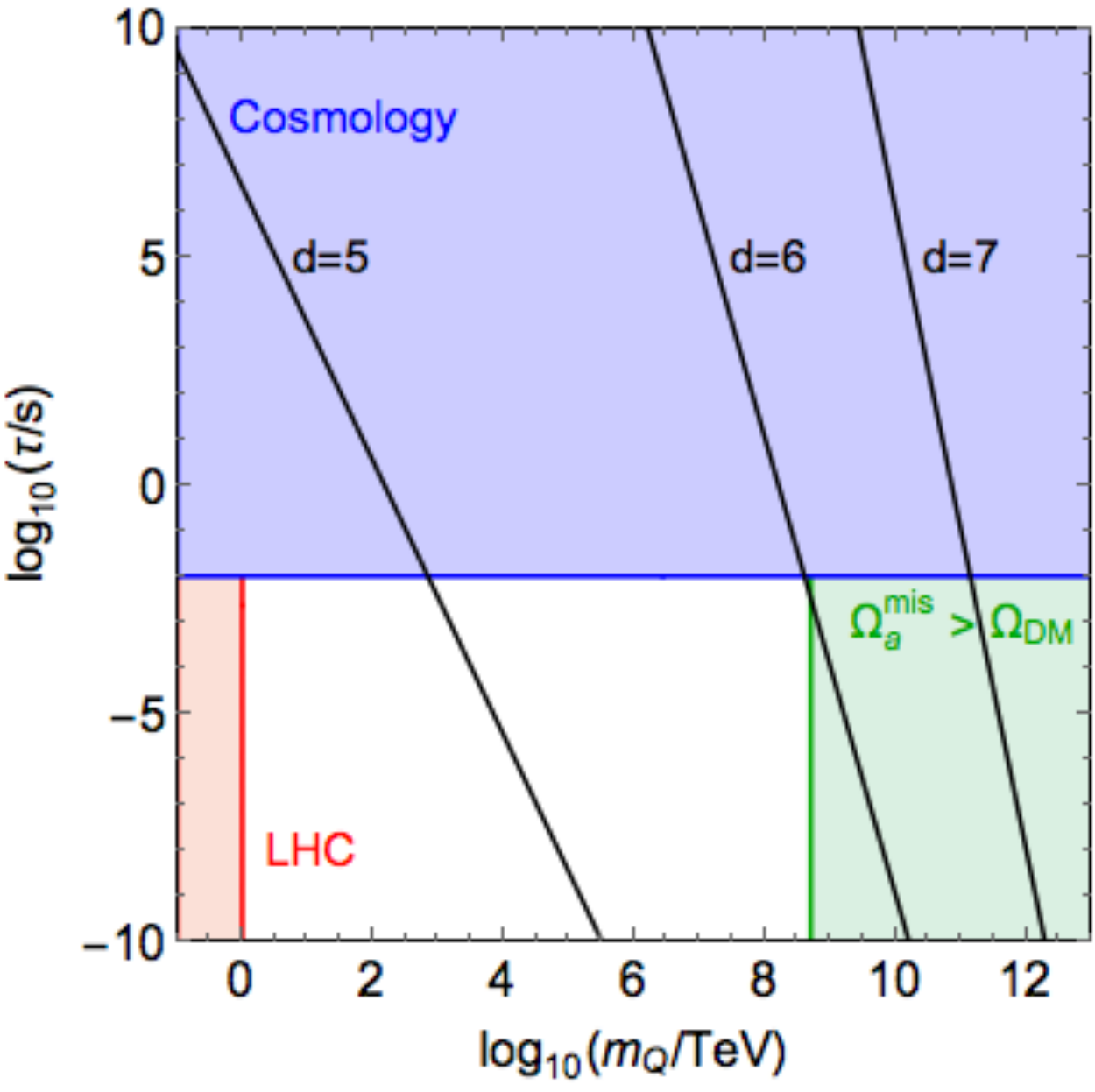}
\caption{\label{fig:lifetimesvsmQ} Heavy quark lifetimes for decays via $d=5,6,7$ 
effective operators. 
The regions in color  are excluded respectively by the BBN limit $\tau_Q
\lsim 10^{-2}\,$s (blue), 
by the LHC limit $m_Q \gsim 1\,$TeV (red), and  by  
requiring $\Omega_a \leq \Omega_{DM}$ which suggests $m_Q \lsim 5\cdot 10^{11}\,$GeV (green). 
}
\end{figure}

Depending on their gauge quantum numbers, the $Q$'s can couple
directly to SM quarks via renormalizable operators.  All the
representations that allow for this possibility are listed
in~\Table{summarydim4}.  We have basically two different cases: $i)$
$d=3$ super-renormalizable operators like, for example,
$\mu_{Qq} \bar Q_L d_R$ as in the first row in~\Table{summarydim4},
or $d=4$ operators involving $\Phi$ which generate effective $d=3$ mixing
operators after PQ symmetry breaking, like for example
$\lambda_{Qq} V_a \bar Q_L d_R$ as in the third row in \Table{summarydim4};
$ii)$ genuine $d=4$ operators, like for example $\lambda_{QqH} \bar q_L Q_R H$
as in the second row in \Table{summarydim4}.  For $m_Q \gsim 10 \,\text{TeV}$,
unless the relevant couplings have exceedingly small values
($\mu_{Qq},\,\lambda_{Qq} V_a \ll 1\,$keV,
$\lambda_{QqH} \ll 10^{-12}$), $\tau_Q < 10^{-2}$ s is always
ensured.
For some $R_Q$'s, even if renormalizable decay operators  are 
forbidden by gauge invariance, effective operators 
of dimension $d > 4$ can still be allowed.  We assume 
conservatively that higher dimensional operators 
are suppressed by powers of the Planck mass $m_P = 1.2 \times 10^{19}$
GeV,
and we write them as: 
\begin{equation}
\mathcal{L}^{d>4}_{Qq} =
\frac{1}{m_P^{(d-4)}} \mathcal{O}_{Qq}^{d>4} + \text{h.c.} \,.  
\end{equation}   
For constant matrix elements and massless final states, the phase space
factor  for $Q$ decays into $n_f$ final states can be integrated
analytically  (see e.g.~\cite{DiLuzio:2015oha}), yielding the decay rate 
\begin{equation}
\label{GammaNDA}
\Gamma_{d,n_f} = \frac{m_Q}{4(4 \pi)^{2 n_f -3}(n_f-1)!(n_f-2)!} \left(\frac{m^2_{Q}}{m^2_P}\right)^{d-4} \!\!\!\! .   
\end{equation}
$Q$-decay operators of $d=5,6,7$ involve at least $n_f=2,3,4$
particles in the final state, thus we obtain:
\hspace{-1cm}
\begin{align}
\tau_{d=5}&= 3.9 \cdot 10^{-20} \ \text{s} \ \left( \frac{5 \cdot 10^{11} \ \rm{GeV}}{m_Q} \right)^3 
\, , \\
\tau_{d=6}&= 7.4 \cdot 10^{-3} \ \text{s} \ \left( \frac{5 \cdot 10^{11} \ \rm{GeV}}{m_Q} \right)^5 
\, , \\
\tau_{d=7}&= 4.2 \cdot 10^{15} \ \text{s} \ \left( \frac{5 \cdot 10^{11} \ \rm{GeV}}{m_Q} \right)^7 
\, ,  
\end{align}
where we have normalized $m_Q$ to its (presumably) largest value
$V_a\lsim 5\cdot 10^{11}\,$GeV compatible, in post-inflationary PQ
breaking scenarios, with an axion energy density not exceeding
$\Omega_{DM}$~\cite{Bonati:2015vqz,Petreczky:2016vrs,%
  Borsanyi:2016ksw}.\footnote{More precisely, the cosmological limit
  holds for $f_a=V_a/N_{DW}$, where $N_{DW}=2N$, so that for
  $N_{DW} >1$ and $y_Q\sim 1$, values $m_Q>5\cdot 10^{11}\,$GeV are also possible,
  opening a small window for the viability of $d=6$ operators.  Since
  this holds only for an ad hoc choice of the couplings, we do not
  include this case among our phenomenologically preferred
  possibilities. }  Our results for $\tau_{d=5,6,7}$ are plotted in
\fig{fig:lifetimesvsmQ}. We see that for $d=5$, decays can occur with
lifetimes shorter than $10^{-2}\,$s as long as $m_Q\gsim 800\,$TeV.
For $d=6$, even when $m_Q \sim V_a$ decays occur dangerously close to
the BBN era. Finally, decays via $d=7$ operators are always excluded.
We can then conclude that only $d\leq 4$ and $d=5$ operators naturally
yield sufficiently fast decays, so that only the $R_Q$ listed in
tables~\Table{summarydim4} and~\Table{summarydim5ALL} are safe from
cosmological issues.

\section{Selection criteria}
\label{axionwindow}

In this section we proceed to select KSVZ-type (or hadronic) axion
models which satisfy the following two criteria: $i)$ cosmologically
safe $Q$ lifetimes, and $ii)$ the absence of LP in the SM gauge
couplings at sub-Planckian scales. We will define as {\it
  phenomenologically preferred} those post-inflationary models
which satisfy these two criteria.  We will also briefly comment on two
other possible criteria, namely the absence of the domain wall (DW)
problem, and $R_Q$-assisted improved gauge coupling unification.
However, we will eventually conclude that these two additional
conditions do not match a sufficient level of generality to represent
reliable selection criteria, and should be better considered just as
desirable features of specific models. After discussing the case of
hadronic (KSVZ) axions in post-inflationary scenarios, in
  \sect{preinflationary} we will generalize the analysis to include 
  pre-inflationary models.


\subsection{First criterium: $Q$ lifetimes}

As a first discriminating criterium we assume that:\ {\it  Models
  that allow for sufficiently short lifetimes
  ($\tau_Q \lsim 10^{-2}\,$s) are phenomenologically preferred with
  respect to models containing long-lived or cosmologically stable
  $Q$'s}. 

According to the analysis in \sect{Qlifetimes}, only
$d\leq 4$ and $d=5$ operators are safe from cosmological issues. 
The quantum number assignments that allow for 
$d\leq4$ decay operators ($\mathcal{L}_{Qq}\neq 0$) 
are collected in \Table{summarydim4}. 
\begin{table}[t!]
\renewcommand{\arraystretch}{1.4}
\centering
\begin{tabular}{@{} |c|c|c|@{}}
\hline
$R_Q$ & $\mathcal{O}_{Qq}^{d\le 4}$ &  $(\mathcal{X}_L,\mathcal{X}_R)$
\\ 
\hline
\hline
\multirow{3}{*}{$(3,1,-1/3)$} 
  & $\overline{Q}_L d_R$ & $(0,-1)$
\\
\cline{2-3}
   & $\overline{q}_L Q_R H$, $\overline{Q}_L d_R \Phi$ & $(1,0)$
\\
\cline{2-3}
  & $\overline{Q}_L d_R \Phi^\dag$ & $(-1,-2)$
\\
\hline
\multirow{3}{*}{$(3,1,2/3)$}  & 
$\overline{Q}_L u_R$ & $(0,-1)$
\\
\cline{2-3}
  & 
$\overline{q}_L Q_R H^\dag$,
$\overline{Q}_L u_R \Phi$ & $(1,0)$
\\
\cline{2-3}
  & 
$\overline{Q}_L u_R \Phi^\dag$ & $(-1,-2)$
\\
\hline
\multirow{3}{*}{$(3,2,1/6)$}  & $\overline{q}_L Q_R$ & $(1,0)$
\\
\cline{2-3}  & $\overline{Q}_L d_R H$, $\overline{Q}_L u_R H^\dag$,  $\overline{q}_L Q_R \Phi$ & $(0,-1)$
\\
\cline{2-3}  & $\overline{q}_L Q_R \Phi^\dag$ & $(2,1)$
\\
\hline
$(3,2,-5/6)$  & 
$\overline{Q}_L d_R H^\dag$ & $(0,-1)$
\\
\hline
$(3,2,7/6)$  & 
$\overline{Q}_L u_R H$ & $(0,-1)$
 \\
\hline
$(3,3,-1/3)$  & 
$\overline{q}_L Q_R H^\dag$ & $(1,0)$
\\
\hline
$(3,3,2/3)$  & 
$\overline{q}_L Q_R H$ & $(1,0)$
\\
\hline
  \end{tabular}
  \caption{\label{summarydim4} 
    $Q$ representations which allow for
    renormalizable couplings  with the SM quarks. 
    The PQ charges $\mathcal{X}_{L,R}$ in the third column are normalized to $\mX_\Phi=1$.
  }
\end{table}
These operators induce 2-body decays either directly or via $Q$-$q$
mass mixing, allowing in both cases for fast $Q$ decays. 
Out of the seven possibilities listed in \Table{summarydim4}, the
ones in the third and fifth row were already identified in
Ref.~\cite{Ringwald:2015dsf}
(see also~\cite{Berezhiani:1989fp,Ballesteros:2016euj,Ballesteros:2016xej}).
 They coincide in fact with models
KSVZ-II and KSVZ-III of~\cite{Ringwald:2015dsf} modulo a redefinition
of the PQ charges by a shift proportional to baryon number
$\mX \to \mX+B'$, respectively with $B'=-\frac{3}{2}$ and
$B'=+\frac{1}{2}$. As long as only $B'$ conserving operators are
considered, this gives no difference in the physics.

In \Table{summarydim5ALL} we give the list of $R_Q$ for which $d=5$
operators involving a single $Q$-insertion are allowed.  In this case
$Q$ decays occur with sufficiently short lifetimes only for
$m_Q\gsim 800\,$TeV.
\begin{table}[ht!] 
\renewcommand{\arraystretch}{1.4}
\centering
\begin{tabular}{@{} |c|c| @{}}
\hline
$R_Q$ &  $\mathcal{O}_{Qq}^{d=5}$  \\ 
\hline
\hline
$(3,3,-4/3)$ & 
$\overline{Q}_L d_R H^{\dag 2}$  \\
\hline
\rowcolor{piggypink}
$(3,3,5/3)$ & 
$\overline{Q}_L u_R H^2$ \\
\hline
\rowcolor{piggypink}
$(3,4,1/6)$ & 
$\overline{Q}_R q_L H^\dag H$, 
$\overline{Q}_R \sigma_{\mu\nu} q_L W^{\mu\nu}$  \\
\hline
\rowcolor{piggypink}
$(3,4,-5/6)$ & 
$\overline{Q}_R q_L H^{\dag 2}$  \\
\hline
\rowcolor{piggypink}
$(3,4,7/6)$ & 
$\overline{Q}_R q_L H^2$  \\
\hline
$(\bar 6,1,-1/3)$ & 
$\overline{Q}_L \sigma_{\mu\nu} d_R G^{\mu\nu}$  \\
\hline
$(\bar 6,1,2/3)$ & 
$\overline{Q}_L \sigma_{\mu\nu} u_R G^{\mu\nu}$ \\
\hline
$(\bar 6,2,1/6)$ & 
$\overline{Q}_R \sigma_{\mu\nu} q_L G^{\mu\nu}$ \\
\hline
$(8,1,-1)$ & 
$\overline{Q}_L \sigma_{\mu\nu} e_R G^{\mu\nu}$  \\
\hline
$(8,2,-1/2)$ & 
$\overline{Q}_R \sigma_{\mu\nu} \ell_L G^{\mu\nu}$  \\
\hline
$(15,1,-1/3)$ & 
$\overline{Q}_L \sigma_{\mu\nu} d_R G^{\mu\nu}$\\
\hline
$(15,1,2/3)$ & 
$\overline{Q}_L \sigma_{\mu\nu} u_R G^{\mu\nu}$  \\
\hline
\rowcolor{piggypink}
$(15,2,1/6)$ & 
$\overline{Q}_R \sigma_{\mu\nu} q_L G^{\mu\nu}$  \\
  \hline
  \end{tabular}
  \caption{\label{summarydim5ALL} 
    $Q$ representations which allow for $d=5$ decay operators.
    $R_Q$ highlighted in red are theoretically disfavored by the  
    appearance of LP at  sub-Planckian energies. 
  }
\end{table}

\subsection{Second criterium: Landau poles}

Large representations can often induce LP in the hypercharge, weak, or
strong gauge couplings $g_1,\, g_2,\,g_3$ at some uncomfortably
low-energy scale $\Lambda_{LP} < m_P$.  At energy scales approaching
$m_P$, gravitational corrections to the running of the gauge couplings
can become relevant, and explicit computations show that they go in
the direction of delaying the emergence of LP~\cite{Robinson:2005fj}.
Therefore, to be conservative, we choose a value of $\Lambda_{LP}$ for
which gravitational corrections are presumably negligible.  As a
second discriminating criterium we then assume that:\ {\it Models in
  which LP in the SM gauge couplings appear only above
  $\Lambda_{LP} \sim 10^{18}\,${\rm GeV} are phenomenologically
  preferred.}  We evaluate the scale at which the LP arise by setting
conservatively the threshold for the $R_Q$ representations at
$m_Q = 5 \cdot 10^{11}$ GeV. In first approximation the scaling of the
LP is linear with $m_Q$, though sizable deviations from linearity are
expected in case of several decades of running.  We employ two-loop
beta functions to avoid possible accidental cancellations which can
arise for some representations in the one-loop
coefficients~\cite{DiLuzio:2015oha,DiLuzio:2015dfa}.  Among the $R_Q$
which allow for $d=5$ decay operators, the five highlighted in light
red in~\Table{summarydim5ALL} lead to LP below $10^{18}$ GeV and we
consider them as theoretically disfavored.

\subsection{Domain walls}

The axion field $a$, being an angular variable, takes values in the
interval $[0, 2\pi V_a)$.  The QCD induced axion potential is periodic
in $a$ with period $\Delta a = 2 \pi V_a / (2N)$, and is thus
characterized by an exact $Z_{2N}$ discrete symmetry.  Once at $T\sim
\Lambda_{QCD}$ the explicit $U(1)_{PQ}$ breaking from non-perturbative
QCD effects starts lifting the axion potential, within the domain of
definition of $a$, $N_{\rm DW}=2N$ degenerate vacua appear, and if the
initial value of $a$ is different in different patches of the Universe
out of causal contact 
(as is the case in post-inflationary scenarios), in each of these
patches the axion field will flow towards a different minimum,
breaking spontaneously $Z_{N_{DW}}$ with $N_{DW}$ different vacuum
values of $a$.  Then, at $T\lsim \Lambda_{QCD}$ DWs will form at the
boundaries between regions of different vacua.  This leads to the
so-called cosmological DW problem~\cite{Sikivie:1982qv} which consists
in the fact that the energy density of the DWs will largely overshoot
the critical density of the Universe.

In pre-inflationary scenarios the problem is avoided at once since the
whole observable Universe corresponds to an initial patch
characterized by a unique value of $a$, and which gets exponentially
inflated to super-horizon scales.  For post-inflationary scenarios
some solutions also exist~\cite{Kim:1986ax,Barr:2014vva}.  The DW
problem is avoided if $N_{DW}=1$~\cite{Vilenkin:1982ks,Barr:1986hs} so
we might want to consider this specific value as an additional
desirable feature for axion models.  In the last column in
\Table{summarydim5} we list the number of DW for each $R_Q$, and we
see that only the first two cases have $N_{DW}=1$.  However, if
multiple $R_Q$ are considered, one can conceive new solutions.  For
instance, given the color indices $T(8)=3$ and $T(6)=5/2$, $N_{DW}=1$
models can be constructed by combining one $(8,1,\mathcal{Y}')$ with a
$(\bar 6,1,\mathcal{Y})$ with opposite PQ charge difference.  With
three $R_Q$ it is also possible to have $N_{DW}=1$ in a trivial way:
by canceling the DW contribution of two $R_Q$ and leaving a third one
with $N_{DW}=1$.

Models with $N_{DW}>1$ can also remain viable in post-inflationary
scenarios, but additional assumptions are needed.  The DW problem can
be disposed of in a simple way by introducing an explicit small soft
breaking of the PQ symmetry~\cite{Sikivie:1982qv}.  On one hand, the
size of this breaking should be large enough so that a single (true)
vacuum can take over before the DWs start dominating the energy
density.  On the other hand, it should be sufficiently small to ensure
that the PQ solution does not get spoiled, as it would occur if
$\theta$ gets shifted away from zero by more than $\sim 10^{-10}$.
Since there is a sizable region in parameter space where these two
conditions can be simultaneously matched~\cite{Marsh:2015xka}, we can
conclude that the DW problem can be solved also in $N_{DW}>1$ models
and, accordingly, we prefer not to consider $N_{DW}=1$ as sufficiently
motivated condition for selecting post-inflationary preferred axion
models.

\subsection{$Q$-assisted unification}

Fixing the threshold for the new quark representations $R_Q$ at some
suitable intermediate scale could improve on SM gauge coupling
unification (see Ref.~\cite{Giudice:2012zp} for a dedicated analysis).
Out of all the representations listed in \Table{summarydim5}, we find
that only $Q \sim (3,2,1/6)$ can considerably improve unification with
respect to the SM while, at the same time, keeping the unification
point at a reasonably high scale
$\Lambda_{\rm GUT} \sim 10^{15}\,$GeV.  This possibility was already
pointed out in~\cite{Giudice:2012zp}, the only difference is that in
our two-loop analysis the value of the optimal threshold
$m_Q = 2 \times 10^{7}$ GeV is about a factor of twenty larger than
what found in~\cite{Giudice:2012zp}.

While gauge coupling unification is a desirable feature for any
particle physics model, envisaging a GUT completion of KSVZ axion
models featuring a hierarchy $V_a \ll \Lambda_{\rm GUT}$ in which only
the fragment $R_Q$ of a complete GUT multiplet receives a mass
$m_Q \lsim V_a\ll \Lambda_{\rm GUT}$, while all the other fragments
acquire GUT-scale masses, is not straightforward. This appears
especially challenging in all the cases in which $U(1)_{PQ}$ commutes
with the GUT gauge group.  Besides these theoretical considerations,
we must also consider the possibility that improved gauge coupling
unification might simply occur as an accident because of the many
representations we have considered.  We then conclude that also
improved unification is not a sufficiently well motivated condition to
be chosen as a selection criterium for preferred axion models.

\subsection{Summary}

The fifteen $R_Q$'s that satisfy our two criteria are collected in
\Table{summarydim5}.  In this table we also give, for each $R_Q$, in
the third column the energy scale $\Lambda_{LP}$ at which the first LP
occurs, together with the corresponding gauge coupling, in the fourth
column the value of $E/N$ which determines the strength of the
axion-photon coupling, and in the last column the number of DW.
\begin{table}[ht!] 
\renewcommand{\arraystretch}{1.2}
\centering
\begin{tabular}{@{} |l|c|c|c|c| @{}}
\hline
$\ \ \ \ \ \ \ \ R_Q$ &  $\mathcal{O}_{Qq}$ & $\Lambda^{\!R_Q}_{LP}$[GeV] & $E/N$ & $N_{\rm DW}$ \\ 
\hline
\hline
$R_1$:$\,(3,1,-\tfrac{1}{3})$ & 
$\overline{Q}_L d_R$ 
& $9.3 \cdot 10^{38} (g_1)$ & $2/3$ & $1$ \\ 
\hline
$R_2$:$\,(3,1,+\tfrac{2}{3})$ & 
$\overline{Q}_L u_R$
& $5.4 \cdot 10^{34} (g_1)$ & $8/3$ & $1$ \\ 
\hline
$R_3$:$\,(3,2,+\tfrac{1}{6})$ & 
$\overline{Q}_R q_L$
& $6.5 \cdot 10^{39} (g_1)$ & $5/3$ & $2$ \\ 
\hline
$R_4$:$\,(3,2,-\tfrac{5}{6})$ & 
$\overline{Q}_L d_R H^\dag$
& $4.3 \cdot 10^{27} (g_1)$ & $17/3$ & $2$ \\
\hline
$R_5$:$\,(3,2,+\tfrac{7}{6})$ & 
$\overline{Q}_L u_R H$
& $5.6 \cdot 10^{22} (g_1)$ & $29/3$ & $2$ \\
\hline
$R_6$:$\,(3,3,-\tfrac{1}{3})$ & 
$\overline{Q}_R q_L H^\dag$ 
& $5.1 \cdot 10^{30} (g_2)$ & $14/3$ & $3$ \\
\hline
$R_7$:$\,(3,3,+\tfrac{2}{3})$ & 
$\overline{Q}_R q_L H$ 
& $6.6 \cdot 10^{27} (g_2)$ & $20/3$ & $3$ \\
\hlineB{2.5}
$R_8$:$\,(3,3,-\tfrac{4}{3})$ & 
$\overline{Q}_L d_R H^{\dag 2}$ & $3.5 \cdot 10^{18} (g_1)$ & $44/3$ & $3$ \\
\hline
$R_9$:$\,(\bar 6,1,-\tfrac{1}{3})$ & 
$\overline{Q}_L \sigma  d_R \cdot G $ & $2.3 \cdot 10^{37} (g_1)$ & $4/15$ & $5$ \\
\hline
$R_{10}$:$\,(\bar 6,1,+\tfrac{2}{3})$ & 
$\overline{Q}_L \sigma  u_R \cdot G $ & $5.1 \cdot 10^{30} (g_1)$ & $16/15$ & $5$ \\
\hline
$R_{11}$:$\,(\bar 6,2,+\tfrac{1}{6})$ & 
$\overline{Q}_R \sigma  q_L \cdot G $ & $7.3 \cdot 10^{38} (g_1)$ & $2/3$ & $10$ \\
\hline
$R_{12}$:$\,(8,1,-1)$ & 
$\overline{Q}_L \sigma  e_R \cdot G $ & $7.6 \cdot 10^{22} (g_1)$ & $8/3$ & $6$ \\
\hline
$R_{13}$:$\,(8,2,-\tfrac{1}{2})$ & 
$\overline{Q}_R \sigma  \ell_L \cdot G $ & $6.7 \cdot 10^{27} (g_1)$ & $4/3$ & $12$ \\
\hline
$R_{14}$:$\,(15,1,-\tfrac{1}{3})$ & 
$\overline{Q}_L \sigma  d_R \cdot G $ & $8.3 \cdot 10^{21} (g_3)$ & $1/6$ & $20$ \\
\hline
$R_{15}$:$\,(15,1,+\tfrac{2}{3})$ & 
$\overline{Q}_L \sigma  u_R \cdot G $ & $7.6 \cdot 10^{21} (g_3)$ & $2/3$ & $20$ \\
\hline
  \end{tabular}
  \caption{\label{summarydim5} 
    $R_Q$ allowing for $d\leq 4$ and $d=5$ decay operators 
    ($\sigma\cdot G \equiv \sigma_{\mu\nu} G^{\mu\nu}$) and yielding 
    LP above $10^{18}$GeV. The scale at which the LP arise is given in the
    third column with, in parenthesis,  the corresponding gauge coupling. 
    The fourth    column lists 
    the anomaly contribution to $g_{a\gamma\gamma}$ and the last
    one the number of DW.} 
\end{table}
%
It should be clear that the two criteria on which our selection has
been based should not be understood as strict no-goes, since under
specific conditions models that do not satisfy these conditions can
also be viable.  For example, the first requirement of sufficiently
fast decays for the strongly interacting relics applies only to
scenarios for which $m_Q < T_{\rm{reheating}}$, and there is no
similar issue in pre-inflationary scenarios. However, in
  \sect{preinflationary} we will show that, as long as the threshold
  for integrating out the heavy $Q$ is kept at, or below,
  $m_Q\sim 5\times 10^{11}\,$GeV, the LP condition alone is
  sufficiently constraining that the previous results get only mildly
  relaxed.  
%
%
  If the PQ breaking scale is allowed to lie sizeably above
  $5\times 10^{11}\,$GeV, as it is possible in pre-inflationary
  models, then also the LP condition will progressively diminish its
  strength due to the possibility of increasing correspondingly the
  heavy quark threshold. This however, can be done only at the cost of 
  an increased fine tuning in the initial value of $\theta$, something
  that  can well be considered as theoretically unpleasant. 
%

\begin{figure*}[t]
\includegraphics[width=.65\textwidth]{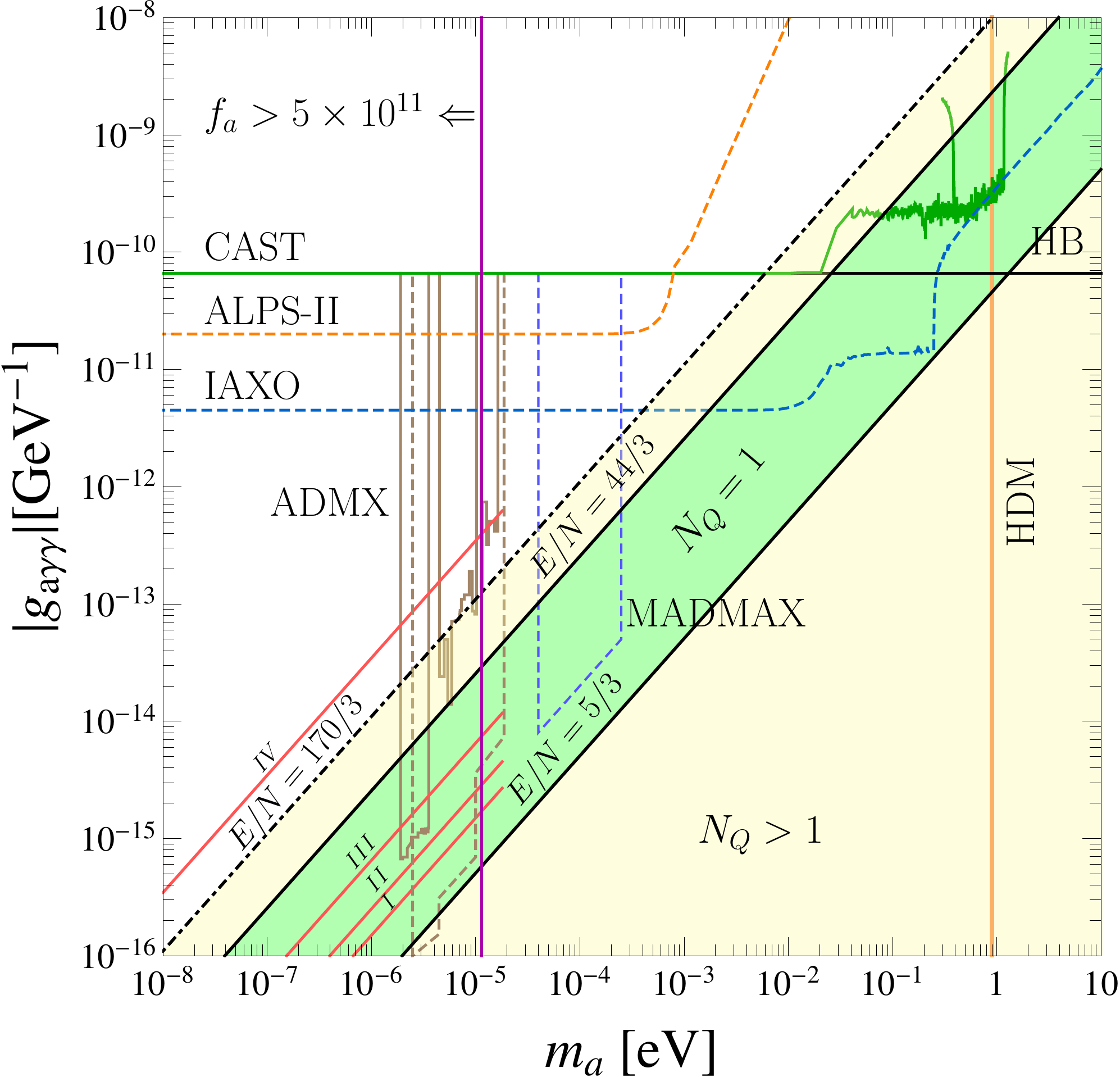}
\caption{\label{fig:Excl}
The $g_{a\gamma\gamma}$-$m_a$ window for preferred 
axion models.  The two lines labeled $E/N=44/3$ and $5/3$  
encompass KSVZ models with a single $R_Q$, while  
the region below $E/N=170/3$  allows for   
more  $R_Q$'s.  The red lines labeled from I to IV
(only partially drawn not to clutter the figure)  
 indicate where  the DFSZ-type of models lie (see \sect{sec:DFSZ}).
Current exclusion regions are 
delimited by  solid  lines. They correspond to the 2017 CAST results
\cite{Anastassopoulos:2017ftl}, to the ADMX limit
\cite{Asztalos:2009yp,Carosi:2013rla,preadmx1,preadmx2,preadmx3}, to
the constraints from hot DM (HDM) \cite{Olive:2016xmw} and from
horizontal branch (HB) stars~\cite{Ayala:2014pea}. The expected
sensitivities for ALPS-II \cite{Bahre:2013ywa}, IAXO
\cite{Carosi:2013rla,Irastorza:2011gs}, ADMX \cite{Asztalos:2003px}
and MADMAX \cite{TheMADMAXWorkingGroup:2016hpc} are depicted with
dashed lines.  On the left hand side of the vertical violet line
labeled $f_a > 5\times 10^{11}\,$GeV the limits for KSVZ models can get
relaxed.  }
\end{figure*}

\section{Axion coupling to photons}
\label{axioncoupling}

From the experimental point of
view, the most promising way to unveil the axion is via its
interaction with photons, which is described by the effective
term 
$\mathcal{L}_{a\gamma\gamma}= - (1/4) g_{a\gamma\gamma} a\, F\cdot
\tilde F$. The axion-photon coupling  is given in terms of the anomaly
coefficients in~\eq{eq:NE} by~\cite{Kaplan:1985dv,Srednicki:1985xd}:
\begin{equation} 
 \label{aphcoupapprox}
g_{a\gamma\gamma}=
\frac{m_a}{\rm{eV}} \ 
\frac{2.0}{10^{10}\ \rm{GeV}} \; \left(\frac{E}{N} - 1.92(4)\right)\, ,
\end{equation}
where the uncertainty is evaluated with the NLO chiral
Lagrangian~\cite{diCortona:2015ldu}.  
The strongest coupling is obtained for
$R^s_Q=(3,3,-4/3)$ that gives $E_s/N_s-1.92\sim 12.75$, almost twice
the usually adopted value of $7.0$~\cite{Olive:2016xmw}, while the
weakest coupling is obtained for $R^w_Q=(3,2,1/6)$ for which
$E_w/N_w-1.92\sim -0.25$ is about 3.5 times larger than the usual
lower value of $0.07$.  
The corresponding couplings are depicted  
in~\fig{fig:Excl} with the two oblique black lines labeled $E/N= 44/3$
and $E/N=5/3$.  According to our two selection criteria all preferred
hadronic axion models containing a single $R_Q$ fall within the light
green strip enclosed by these two lines.

Let us now study to which extent the previous results can be changed
by the presence of more $R_Q$'s. It would be quite interesting if, for
example, $g_{a\gamma\gamma}$ could receive sizeable
enhancements. However, we can easily see that, as long as the sign of
$\Delta\mX = \mX_L-\mX_R$ is the same for all $R_Q$'s, this cannot
occur. Let us write the combined anomaly factor for $R_Q + R^s_Q$:
\beq 
\frac{E_c}{N_c}\equiv 
\frac{E+E_s}{N+N_s} = \frac{E_s}{N_s} \left(\frac{1 +
    E/E_s}{1+N/N_s}\right) \, . 
\eeq 
Since by construction the anomaly coefficients of any $R_Q$ in our
preferred set satisfy $E/N\leq E_s/N_s $, the factor in parenthesis
cannot be larger than unity, implying $E_c/N_c < E_s/N_s$.  This is
not so, however, if we allow for opposite signs in the PQ charge
differences: $\Delta\mX= - \Delta\mX^s$.  In this case $E/E_s$ and
$N/N_s$ become negative and $g_{a\gamma\gamma}$ can get enhanced.  The
largest enhancement attainable with two $R_Q$'s is obtained with
$R_Q^s \oplus R_Q^w$. This still respects the LP selection criterium
and yields $E_c/N_c = 122/3$.  Further enhancements are possible with
three or even more $R_Q$'s, but adding multiple $R_Q$'s will
eventually lead to sub-Planckian LP, so that there is in fact an
absolute upper bound on $E_c/N_c$ compatible with the requirement of
no LP below $10^{18}$ GeV. We have searched for this maximum coupling,
and we have found that it corresponds to the combination
$R_{8} \oplus R_{6} \ominus R_{9}$, where the meaning of the symbols
$\oplus$ and $\ominus$ is that the representation has to be taken with
the same or with the opposite sign of the PQ charge difference
$\Delta \mX$.  The resulting maximum axion-photon coupling corresponds
to $E/N= 170/3$ and is depicted in~\fig{fig:Excl} with the uppermost
dot-dashed oblique line.

Besides enhancing the axion-photon coupling, more $R_Q$'s can also
weaken $g_{a\gamma\gamma}$ below the lower limit $E/N=5/3$ for a
single $R_Q$, and even yield complete axion-photon decoupling (within
theoretical errors), a possibility that requires an ad hoc choice of
$R_Q$'s, but no numerical fine tuning.  With two $R_Q$'s there are
three such cases: $R_6 \oplus R_9$; $R_{10} \oplus R_{12}$ and
$R_4 \oplus R_{13}$ giving respectively
$E_c/N_c = (23/12, 64/33,41/21) \approx (1.92,1.94,1.95)$.  In all
these cases the axion could be detected more easily via its coupling
to nucleons, providing additional motivations for axion searches which
do not rely on the axion coupling to
photons~\cite{Budker:2013hfa,Arvanitaki:2014dfa,Barbieri:2016vwg}.\footnote{Note
  that in KSVZ-type of axion models the coupling to nucleons is
  model-independent (see e.g.~\cite{diCortona:2015ldu}), while the
  axion coupling to electrons is loop suppressed
  \cite{Srednicki:1985xd}.}  Finally, let us mention that in the cases
in which $g_{a\gamma\gamma}$ is strongly suppressed, the limits from
stellar evolution are accordingly weakened.  However, the region
$m_a \gtrsim 1\,$eV is still excluded due to the hot DM bound
\cite{Olive:2016xmw}.
 
\section{KSVZ models in pre-inflationary scenarios}
\label{preinflationary}

The discussion of the KSVZ models in the previous sections pertained
to post-inflationary scenarios, with $m_Q < T_{\rm reheating}$
ensuring, as initial condition, a thermal abundance for the $Q$.
However, the scale of inflation is firmly bounded from below only by
BBN considerations, which imply a loose limit of just a few MeV, and
thus $m_Q \gg T_{\rm reheating}$ is certainly not an unlikely
possibility.  It is then mandatory to explore to which extent our
results can represent acceptable estimates of the preferred axion
window also in pre-inflationary scenarios, when the first condition of
forbidding long lived or stable strongly interacting relics must 
be dropped.

The requirement that the contribution to the cosmological energy
density from axion misalignment does not exceed the energy density of
DM implies, in post-inflationary scenarios,
$f_a \lsim 5\cdot 10^{11}\,$GeV.  In pre-inflationary scenario this
condition can be avoided by assuming that in the original patch
corresponding to the present observable Universe the initial value of
$\theta$ is sufficiently close to the minimum of the zero temperature
axion potential.  However, values largely in excess of
$f_a \sim 5\cdot 10^{11}\,$GeV require correspondingly large fine
tunings in the initial conditions, or invoking anthropic selection
arguments to justify a sufficiently small initial value of
$\theta$. This might well be considered an unwanted feature of {\it
  preferred axion models}, and therefore we will restrict our study of
the pre-inflationary case by keeping the condition
$f_a \lsim 5\cdot 10^{11}\,$GeV.  Taking also in this case a maximum
value for the Yukawa couplings $y_Q\leq 1$, we can again apply the LP
criterium with $m_Q = 5\cdot 10^{11}\,$GeV as the threshold for
integrating out the heavy quarks.

To see which constraints can be implied by the LP condition alone, let us
start by considering a single representation
$R_Q\!\!=\!  (\mathcal{C}_Q,\mathcal{I}_Q,\mathcal{Y}_Q)$. The $E/N$
factor can be conveniently written as
\beq
\label{EovNgeneralexp}
\frac{E}{N} = \frac{\mathcal{C}_Q}{T(\mathcal{C}_Q)} 
\left[ \frac{1}{12} (\mathcal{I}_Q^2
    -1) + \mathcal{Y}_Q^2 \right] \, , 
\eeq
where $\mathcal{C}_Q$ is the dimension of the color representation, 
and  $T(\mathcal{C}_Q)$ is the color Dynkin index previously
introduced. In terms of the Dynkin indices 
labeling the representation $ \mathcal{C}_Q= (\alpha_1,\alpha_2)$
the index can be written as~\cite{Slansky:1981yr}:
\begin{eqnarray}
\label{eq:index}
T(\mathcal{C}_Q) &=&\frac{1}{24} \mathcal{C}_Q P(\alpha_1,\alpha_2)\,,
  \\
P(\alpha_1,\alpha_2) &=& \left(\alpha_1^2 + 3 \alpha_1 +\alpha_1 \alpha_2
  +3\alpha_2+\alpha^2_2\right).\quad  
\end{eqnarray}
The polynomial $P(\alpha_1,\alpha_2)$ which appears in the denominator
of $E/N$ has its minimum value for the fundamental representation
$(\alpha_1,\alpha_2)=(1,0)$, so we learn that the largest values of
$E/N$ are obtained for a color triplet.

In order to study the values of $E/N$ in 
$(\mathcal{I}_Q,\mathcal{Y}_Q)$ representation space, we start by
establishing the `corners' that saturate the LP condition.
Respectively for hypercharge and weak isospin we find:
\begin{eqnarray}
  \label{eq:corner1}
  R_Q(\mathcal{Y}_Q^{\rm max}) &=& (3,1,5/2) \ \longrightarrow \ E/N=75/2\, ,\quad \\
  \label{eq:corner2} 
  R_Q(\mathcal{I}_Q^{\rm max}) &=& (3,4,0) \quad\ \longrightarrow \ E/N=15/2\,. 
\end{eqnarray}
Any larger value of hypercharge or isospin would induce a
sub-Planckian LP in the corresponding coupling (although in our search
we allow for continuous values of $\mathcal{Y}_Q$, for convenience we
round the result to a close fractional value). We can now use
\eqn{EovNgeneralexp} to find the maximum of $E/N$ subject to the
condition $\mathcal{Y}_Q \sqrt{\mathcal{I}_Q} \leq \frac{5}{2}$, 
which is implied by the maximum allowed coefficient for the
hypercharge coupling $\beta$-function. 
This value  is given by the value of $\mathcal{I}_Q$ that maximizes the
function
\begin{equation}
  \label{eq:maxLP}
F(\mathcal{I}_Q) =  \mathcal{I}_Q^2 - 1 +\frac{75}{\mathcal{I}_Q}
\end{equation}
within the allowed domain.  $F(\mathcal{I}_Q)$ is a parabola-shaped
function with a minimum in $\mathcal{I}_Q=(75/2)^{1/3}\approx 3.35$.
The value of $E/N$ for $\mathcal{I}_Q=1$ is approximately matched only
at $\mathcal{I}_Q=8$, which is much larger than the LP constraint on
the maximum dimension of weak-isospin representation
($\mathcal{I}_Q =4$). Thus, in the case of a single representation,
$R_Q=(3,1,5/2)$ in \eqn{eq:corner1} gives the maximum value compatible
with the LP condition: $E/N=75/2$.  With respect to the results
obtained in the post-inflationary case, this is 2.5 times larger than
the upper bound for a single $R_Q$ ($E/N = 44/3$), but still much
smaller than the upper bound obtained by allowing for more $R_Q$
($E/N = 170/3$).

In the case of more representations, the maximum value of $E/N$ can be
found in correspondence of the `corners' of the
$(\mathcal{I}_Q,\mathcal{Y}_Q)$ representation space (as well as for
sets of representations that are `equivalent' in the sense specified
below).  The combination $(3,1,5/2)\oplus(3,4,0)$ which maximizes the
numerator of $E/N$ requires the addition of $SU(2)_L\times U(1)_Y$ singlet
and color non-trivial representations, in order to minimize the
denominator to the minimum possible value of $\pm 1/2$. With three
representations, adding $\ominus(8,1,0)$ (with a negative sign of the
PQ charge difference) allows to arrange for this given that
$(1+4)\cdot T(3)-T(8)=-1/2$.  For this combination of three $R_Q$ we
then obtain $E/N= - 135/2$. With four representations, the $+1/2$ in
the denominator can be obtained by including
$ \oplus(3,1,0)\ominus(6,1,0)$. This gives $E/N= + 135/2$ which
results in a slightly smaller axion-photon coupling, due to the
negative sign of the chiral perturbation contribution, see
\eqn{aphcoupapprox}.  Equivalent representations can be obtained for
example by the replacement
$(3,1,5/2) \oplus(3,1,0) \to (3,1,\mathcal{Y}_Q) \oplus
(3,1,\mathcal{Y}_{Q'})$
with $\mathcal{Y}_Q^2+\mathcal{Y}_{Q'}^2=(5/2)^2$, as well as in other
similar ways.  To check that the result obtained relying on the
previous simple argument indeed corresponds to the maximum $(E/N)$, we
have carried out a thorough computer search exploring
$(\mathcal{I}_Q,\mathcal{Y}_Q)$ representation space, which confirmed
that $|E/N| = 135/2$ gives the maximum axion-photon coupling
compatible with the LP condition. The corresponding upper limit on
$g_{a\gamma\gamma}$ is only about 20\% larger than the maximum
coupling labeled $E/N = 170/3$ obtained in post-inflationary
scenarios, and well below the limit on DFSZ-IV models represented by
the uppermost red line. Not to clutter too much the plot, we have not
included in \fig{fig:Excl} the corresponding line.

In conclusion, for KSVZ models the preferred axion window for the
different cases considered is well represented by the black oblique
lines in \fig{fig:Excl}, restricted to the region on the right hand
side of the violet vertical line labeled $f_a > 5\times 10^{11}\,$GeV.
On the left of this line only pre-inflationary models with
progressively larger values of $f_a$ are allowed.  In this case the
heavy quark threshold can be correspondingly increased, thus weakening
the constraints from the LP condition.  Therefore for
KSVZ models larger values of the axion-photon coupling become allowed
within this region. However, this goes at the expense of a
progressively larger amount of fine tuning in the initial value of
$\theta$, which might well be considered as an unwanted feature in
phenomenologically preferred axion models.

\section{DFSZ-type of axion models}
\label{sec:DFSZ}

In DFSZ-type of models~\cite{Zhitnitsky:1980tq,Dine:1981rt} two or
more Higgs doublets $H_i$, carrying PQ charges, together with the SM
singlet axion field $\Phi$ are introduced. The SM fermion content is
not enlarged, but in general both quarks and leptons carry PQ
charges.  The electromagnetic and color $U(1)_{PQ}$
anomalies then depend on the known fermions assignments under the SM gauge
group, but also on their model dependent PQ charge assignments.
Hence, several variants of DFSZ axion models are possible, some of
which have been discussed, for instance, in
Refs.~\cite{Cheng:1995fd,Kim:1998va}.  Here we argue that for most of
these variants the axion-photon coupling falls within the regions 
highlighted in Fig.~\ref{fig:Excl}. Only in some specific cases the
KSVZ upper limit $E/N=170/3$ can be exceeded. We will point out under
which conditions this can occur.

Let us start with some general considerations: we assume $n_H\geq 2$
Higgs doublets $H_i$ which are coupled to quarks and leptons
via Yukawa interactions, and to the axion field $\Phi$ through scalar
potential terms.  The kinetic term for the scalars carries a
$U(1)^{n_H+1}$ rephasing symmetry that must be explicitly broken to
$U(1)_{PQ}\times U(1)_Y$ in order that the PQ current in \eq{eq:NE} is
unambiguously defined, and to avoid additional Goldstone bosons with
couplings only suppressed as the inverse of the electroweak scale.  By considering
from the start only gauge invariant operators, the relevant explicit
breaking $U(1)^{n_H+1}\to U(1)_{PQ}$ must be provided by non-Hermitian
renormalizable terms in the scalar potential involving $H_i$ and
$\Phi$.  This implies that the PQ charges of all the fermions and
Higgs doublets are interrelated and cannot be chosen arbitrarily.  In
the most general scenario, each SM fermion field carries a specific PQ
charge. However, given that the anomalies of the PQ current depend on
the difference between the PQ charges of L- and R-handed 
fermions, without loss of generality we can set the PQ charges of the
L-handed fermions to zero, and only consider the charges 
of the R-handed fermions  
$ \xu j,\, \xd j,\,  \xe j$, 
%
%
%
where $j$ is a generation index.  
The  ratio of anomaly coefficients $E/N$ reads 
\begin{align}
\label{EoNgeneralDFSZ}
\nonumber
\frac{E}{N} &= 
 \frac{\sum_j \left( \frac{4}{3} \xu j  
+ \frac{1}{3} \xd j + \xe j\right)}{\sum_j \left( \frac{1}{2} \xu j +
              \frac{1}{2} \xd j \right)} \\
&= \frac{2}{3} + 2\frac{\sum_j \left( \xu j + \xe j \right)}{
 \sum_j \left( \xu j + \xd j\right)} \, , 
\end{align}
and it is particularly convenient to write it as in the second equality.
Note that in order to have a non-vanishing PQ-color anomaly, the
denominator 
must be non-vanishing.
The original DFSZ model~\cite{Zhitnitsky:1980tq,Dine:1981rt} includes
two Higgs doublets, $H_{u,d}$, coupled to the singlet scalar field via
the quartic term $H_u H_d \Phi^2$, and family independent PQ charges
for the SM fermions. Then the factor $E/N$ is fixed up to the two-fold
possibility of coupling the leptons either to $H_d$ or to $H_u^*$.
\eq{EoNgeneralDFSZ} shows that these two cases yield, respectively
\begin{align}
\nonumber
\label{DFSZ-12}
\text{DFSZ-I}:& \qquad \mX_e =\mX_d  \, ,\ \ \  \quad  E/N= 8/3 \, , \\  
\text{DFSZ-II}:& \qquad \mX_e =-\mX_u \, , \quad E/N= 2/3 \, , 
\end{align}
which in both cases give axion-photon couplings that fall inside the
KSVZ band in Fig.~\ref{fig:Excl}.

Let us now consider the so called DFSZ-III variant~\cite{Cheng:1995fd}
in which the scalar sector is enlarged to contain $n_H=3$ Higgs
doublets $H_{e,d,u}$ coupled respectively to leptons, down-type
and up-type quarks.  Although here we have some more freedom in
choosing the values of the charges $\mX_e$, in order to enforce the
breaking
$U(1)^4 = U(1)_e\times U(1)_u\times U(1)_d\times U(1)_\Phi\to
U(1)_{PQ}$,
$H_e$ must couple to $H_u$, $H_d$ and/or $\Phi^2$, so that $\mX_e$
cannot be completely arbitrary. To find the maximum allowed value, let us
consider the bilinear mixed scalar monomials
$(H_e H_u)\,,(H_e^* H_d),\, (H_u H_d)$ together with their Hermitian
conjugates, responsible for $U(1)^4$ breaking.  It is easy to verify
that the bilinear terms alone yield the same two possibilities listed in
\eq{DFSZ-12}.  Let us then consider quadrilinear couplings.
Since $\Phi^2$ has the same PQ charge than $(H_u H_d)^\dagger$, the
four cases below exhaust all the possible relations between $\mX_e$
and the other PQ charges:
\begin{eqnarray}
\label{quadlinear} \nonumber
(H_eH_u)\cdot (H_u H_d) \  \;   &\Longrightarrow&   \mX_e =  -(2\mX_u+\mX_d)\,,
                                                           \qquad \\ \nonumber
(H_e H_u) \cdot (H_u H_d)^\dagger   \;   &\Longrightarrow&  \mX_e =
                                                           \mX_d\,, \\ \nonumber
(H_e^*H_d) \cdot  (H_u H_d)       \ \;   &\Longrightarrow&  \mX_e =
                                                           \mX_u+2
                                                           \mX_d\,, \\ 
(H_e^*H_d) \cdot  (H_u H_d)^\dagger \;  &\Longrightarrow&   \mX_e = -\mX_u\,. 
\end{eqnarray} 
These four possibilities yield, respectively: 
\begin{equation}
\label{eq:DFSZ-III}
\text{DFSZ-III}: \quad\ \  E/N=-4/3,\,8/3,\,14/3,\,2/3 
\end{equation}
all of which give axion-photon couplings that fall
within the $N_Q=1$ band in Fig.~\ref{fig:Excl}.\footnote{Note that
  the $\mX_{e,u,d}$ charges of the DFSZ-III variants in
  Ref.~\cite{Cheng:1995fd} do not allow to build PQ and gauge
  invariant renormalizable mixed terms.  Consequently,
  $U(1)^4$ cannot get broken to a single $U(1)_{PQ}$ or, in other
  words, $\mX_e$ remains unrelated to the PQ symmetry acting on the
  quarks.}

Many more possibilities in choosing the PQ charges become possible if
we allow for generation dependent assignments, as was done for
example in Ref.~\cite{Bertolini:2014aia}.  The maximum freedom
corresponds to the case in which there are three Higgs doublets for
each fermion species ($H_{e_1}, H_{e_2}, H_{e_3},\;$ etc.) so that the
scalar rephasing symmetry is $U(1)^{n_H+1}$ with $n_H=9$. Here, we refer
to this scenario as DFSZ-IV. Although such a model might be plagued by
various phenomenological issues, bounding from above the maximum
possible $E/N$ in DFSZ-IV is useful, since it provides an upper bound
to $E/N$ for all cases with generation dependent PQ charges, and
with $n_H \leq 9$ Higgs doublets coupled to the SM fermions.

From \eq{EoNgeneralDFSZ} we see that in order to maximize $E/N$ we
have to find  the maximum possible value of $\sum_j (\xu{j} + \xe{j})$ (namely
the largest possible PQ charges for the up-type quarks and
leptons, all with the same sign) together with the minimum value of
the denominator $\sum_j (\xu j+\xd j)$ compatible with a nonzero QCD anomaly,
which  is $ 2 \mX_\Phi$. This second condition is
realized, without loss of generality, by choosing
\beq
\xu{1}= 2 \mX_\Phi,\quad  \xd{1}=\sum_{j=2}^3    (\xu j+\xd j)= 0\,.
\eeq
The last equality is satisfied by 
\beqa
\nonumber
\xu 2 = y,  && \qquad \xd 2 = -y + 2\mX_\Phi,\\
\xu 3 = z,  && \qquad \xd 3 = -z - 2\mX_\Phi\,,  
\eeqa
where the values of $y$ and $z$ are arbitrary.  The scalar terms
allowed by this choice break the $U(1)^4$ symmetry in the second and
third generations to $U(1)_y\times U(1)_z$, which in turn must be
broken by couplings between these scalars and scalars of the first
generation.  Starting with the second generation, the term
$H_{d_2} H_{u_1}(\Phi^*)^2$ yields the largest possible charge
$\xu{2}=y=6 \mX_\Phi $ (and $\xd{2}=-4 \mX_\Phi $). Note that the term
$H_{u_2} H_{u_1}^* \Phi^2$ would instead only yield $y=4 \mX_\Phi $.
The relatively large charge $\xu{2}$ allows to get an even larger
charge $\xu{3}$ via the term
$(H_{u_3} H_{u_2}^*)(H_{d_1}^* H_{u_2}^*)$ giving
$\xu{3}=z=12\mX_\Phi$ and $\xd{3}=-14\mX_\Phi$. This accomplishes the
breaking of all the redundant symmetries in the quark sector.
Regarding the breaking of the $U(1)^3$ symmetries in lepton sector, we
need to couple at least one lepton scalar ($H_{e_1}$ without loss of
generality) to the scalars of the quark sector. The possible bilinears
are either of the form $(H_{e_1} H_{u_j})$ or $(H_{e_1} H_{d_j}^*)$.
The most favorable possibility to get a large charge $\xe{1}$ is to
start with $(H_{e_1} H_{d_1}^*)$, since $H_{d_1}$ has the only
non-negative charge $\xd{1}=0$, and next to couple this term to the
bilinear with the largest possible positive charge, which is
$(H_{d_3} H_{d_1}^*)$.  This yields $\xe{1}=14\mX_\Phi$, which can be
used to push up $\xe2$ and $\xe3$ to even larger values, via the
following sequence of couplings:
$(H_{e_2} H_{e_1}^*) (H_{d_3} H_{e_1}^*)$, yielding $\xe{2}=42$, and
$(H_{e_3} H_{e_2}^*) (H_{d_3} H_{e_2}^*)$, yielding $\xe{2}=92$.  The
values of the PQ charges derived in this way give the maximum possible
axion-photon coupling in DFSZ-IV models, which corresponds to
\begin{equation}
\label{eq:DFSZ-IV}
\text{DFSZ-IV}: \quad\ \  \left(E/N\right)_\text{max}= 524/3 \,. 
\end{equation}
In this class of models it is also easy to obtain axion-photon
decoupling ensuring  at the same time a  
correct breaking of the $U(1)^{9+1}$ global
symmetries down to $U(1)_{PQ}$. An  example  is given by:
\beqa
\nonumber
\xu{j}&=&(2,4,8)\,\mX_\Phi\,,\\ \nonumber
\xd{j}&=&(0,-2,-4)\,\mX_\Phi\,,  \quad \\ \nonumber
\xe{j}&=&(-1,-3,-5)\,\mX_\Phi\,,
\eeqa
which yields $E/N=23/12 \approx 1.92$.  In conclusion, although the
value in \eq{eq:DFSZ-IV} exceeds by a factor of three the maximum KSVZ
value $E/N=170/3$, the construction through which $(E/N)_{\text{max}}$
has been obtain is sufficiently cumbersome to suggest that the $N_Q>1$
region in Fig.~\ref{fig:Excl} can be considered as representative also
of most of DFSZ-IV models.  The values of $E/N$ associated to the
maximum and minimum of $g_{a\gamma\gamma}$ for different classes of
models are summarized in \Table{summaryEoN}.  Note that
  differently from the KSVZ models analyzed previously, the limits on
  the axion-photon coupling in DFSZ models do not depend on details of
  the Universe cosmological evolution, and therefore hold also within
  the region on the left of the violet vertical line in \fig{fig:Excl}
  labeled $f_a > 5\times 10^{11}\,$GeV.

\begin{table}[t!] 
\renewcommand{\arraystretch}{1.2}
\centering
\begin{tabular}{@{} |l|c|c| @{}}
\hline
 &  $E/N(g^{\text{max}}_{a\gamma\gamma})$ & 
 $E/N(g^{\text{min}}_{a\gamma\gamma})$  \\ 
\hline
\hline
KSVZ ($N_Q=1$) &  ${44}/{3}$ &  ${5}/{3}$  \\ 
KSVZ ($N_Q>1$) &  ${170}/{3}$ &  ${23}/{12}$  \\ 
\hline
DFSZ-I-II ($n_H=2$) &  ${2}/{3}$ &  ${8}/{3}$  \\ 
DFSZ-III ($n_H=3$) &  $-{4}/{3}$ &  ${8}/{3}$  \\ 
DFSZ-IV ($n_H=9$) &  ${524}/{3}$ &  ${23}/{12}$   \\ 
\hline
  \end{tabular}
  \caption{\label{summaryEoN} 
    Values of $E/N$ corresponding to the maximum and minimum values of
    $g_{a\gamma\gamma}$ for different classes of models.  Only KSVZ models 
    that satisfy both the selection rules discussed in the text are included.} 
\end{table}

\bigskip

 \section{Axion-photon couplings above the axion window} 
\label{sec:clockwork}

As we have seen, the criterium of the absence of LP plus the
requirement of no cosmological issues in post-inflationary scenarios,
or  of natural initial values for $\theta$ in pre-inflationary
scenarios, imply well defined limits for the axion-photon coupling in
all the type of models we have considered so far.  
However, it is also possible to envisage scenarios in which our
selection criteria are satisfied, and still the axion-photon coupling
can lie well above the preferred window.  The models we have
considered in our analysis are characterized by a specific sector of
scalar fields carrying PQ charges.  In KSVZ-type of models we have
included only one SM singlet scalar $\Phi$.  In DFSZ-type of models we
have allowed for up to one scalar doublet for each type of SM fermion,
for a total of nine electroweak doublets carrying PQ charges, in
addition to the singlet $\Phi$.  However, more PQ charged singlets
$\Phi_k$ could be introduced without conflicting with phenomenological
constraints, and up to about fifty electroweak scalar doublets $H_k$ could
be also added before violating the LP condition.  By adding scalar
doublets that do not couple directly to the fermions, it is possible
to obtain very large PQ charges for the leptons, and huge enhancements
in the value of $E/N$.  To see how this can work let us start from the
quadrilinear scalar coupling $H_u H_d \Phi^2$ and the PQ charges
$\mX(H_u) = -2\mX_\Phi$ and $\mX(H_d)=0$.  Let us define $H_1=H_u$ and
next let us add a whole set of scalar doublets $H_k$ ($k=2,3,\dots,n$)
with charges $\mX(H_k) = -2^k \mX_\Phi$, coupled as $(H_{k} H_{k-1}^*)
(H_{k-1}^* H_d^*)$. Finally, let us couple the lepton Higgs doublet as
$(H_e H_n)(H_n H_d)$.  We obtain $\mX(H_e)= 2^{n+1} \mX_\Phi$.  As
mentioned above, $n$ can be as large as fifty before a LP is hit below
the Planck scale, so that exponentially large lepton charges $|\mX_e|
\sim 2^{50}$ are possible (in this construction the axion-electron
coupling gets exponentially enhanced as well). An analogous
construction is possible also in KSVZ models by adding more scalar
singlets $\Phi_k$.  This possibility was put forth
in~\cite{Farina:2016tgd} to which we refer for further details.

\section{Conclusions}
\label{concl}

Axions are well-motivated candidates for physics beyond the SM.  Axion
models solve the strong CP problem and provide an excellent DM
candidate. Most importantly, experiments are starting \emph{now} to
probe the parameter space region for the axion-photon coupling
predicted by realistic axion models, and an outburst of new
experimental proposals for axion searches have been put forth in last
few years (see e.g.~\cite{Budker:2013hfa,Kahn:2016aff,Chung:2016ysi,%
  TheMADMAXWorkingGroup:2016hpc,Rybka:2014cya,Barbieri:2016vwg,
  Arvanitaki:2014dfa}).  It is then very important to define precisely
the region in parameter space where axion searches should focus,
possibly assessing which are the desirable properties common to axion
models that fall within this region.  The commonly adopted axion
window considered so far~\cite{Olive:2016xmw} corresponds to a
selection of realistic KSVZ and DFSZ axion models, e.g.~those of
Refs.~\cite{Kaplan:1985dv,Cheng:1995fd,Kim:1998va}; but, lacking of a
well-defined guiding principle, it unavoidably contains some degree of
arbitrariness.

In this work we have put forth a recipe for defining a window for
preferred axion models on the basis of  well defined sets of 
selection criteria.  We have considered first KSVZ-type of axion
models, for which all the particles carrying PQ charges are new,
beyond the SM states. 
Starting with post-inflationary scenarios we have classified the representations $R_Q$
of the new quarks $Q$ on the basis of the following two criteria:
$(i)$ $Q$ decays must be fast enough in order not to bring in
cosmological issues; $(ii)$ the new representations $R_Q$ should not
generate Landau poles below $10^{18}\,$GeV.  Only fifteen
representations which we have collected in \Table{summarydim5} satisfy
these two selection criteria.  The ratio of their anomaly coefficients
$E/N$ can then be used to define a first window for preferred axion
models, which is displayed in \fig{fig:Excl} in the
$m_a$-$g_{a\gamma\gamma}$ plane.  We have then shown that models
containing multiple $R_Q$ representations, but which still satisfy the
two criteria, allow to enlarge sizeably this window, and that at fixed
values of $m_a$ both stronger and weaker couplings are possible.
While the weakening of the coupling can reach the limit of complete
axion-photon decoupling (within current theoretical errors), the size of the
possible enhancements is still bounded by an upper limit on $E/N$,
which is set by the LP condition.  We have also discussed the
possibility of considering additional criteria, like requiring the
absence of cosmologically dangerous domain walls, or considering the
possible improvements in SM gauge coupling unification induced by
$R_Q$, but we have concluded that these conditions are not
sufficiently strong and general to represent valid selection criteria,
and should just be considered as desirable features of specific
models.

We have then extended the analysis to include also KSVZ models in
pre-inflationary scenarios, in which case the first condition {\it
  (i)} must be dropped.  The second condition {\it (ii)} on the
absence of sub-Planckian LP still holds, and maintains in full its
constraining power under the assumption that the threshold for
integrating out the new heavy quarks remains at, or below,
$m_Q\sim 5\cdot 10^{11}\,$GeV. This corresponds to axion models in
which the QCD $\theta$ parameter can assume natural initial values, of
order unity, without generating an overabundance of DM.  We have shown
that, with respect to the case when also condition {\it (i)} can be
consistently applied, the upper limits on the axion-photon couplings
in pre-inflationary scenarios are relaxed by a factor of 2.5 in the
case of a single $R_Q$, and only by 20\% in the case of multiple
$R_Q$.

Finally, we have argued that the definition of a preferred axion
window based on the analysis of post- and pre-inflationary KSVZ-type
of axion models, is also representative of the vast majority of
realistic DFSZ-type of models.  The minimal DFSZ realization contains
two Higgs doublets, and the axion-photon coupling is fixed up to a
two-fold choice. The next-to-minimal realization includes one
additional Higgs doublet coupled to the leptons. Since leptons
contribute only to the anomaly coefficient $E$, this allows for larger
values of $E/N$.  Nevertheless, for all these cases the axion-photon
coupling falls within the window for post-inflationary KSVZ
models with a single $R_Q$. A much more general (and probably
unrealistic) possibility includes nine Higgs doublets, each coupled to
a different SM fermion.  The maximum $E/N$ allowed in this extreme
case still exceeds the limit for post-inflationary hadronic
axion models with multiple $R_Q$ representations by only a factor of
three.  Sizeable enhancements of the axion-photon couplings seem to
become possible only in rather cumbersome constructions which
introduce a large number of scalars carrying PQ charges (electroweak
singlets for KSVZ models~\cite{Farina:2016tgd} or doublets for DFSZ
models) which do not couple to the SM fermions or to the $Q$, but only
among themselves through a quite specific pattern of mixed operators.

\section*{Acknowledgments}

\noindent
F.M.~acknowledges financial support from FPA2013-46570, 2014-SGR-104
and MDM-2014-0369. E.N.~is supported in part by the INFN ``Iniziativa
Specifica'' TAsP-LNF. L.D.L.~is grateful to the LNF theory group for
hospitality and financial support during the development of this project.

\appendix 

\section{Fractionally charged $Q$-hadrons and  $Q$ stability} 
\label{app:integer-charged}

In this appendix we argue that $R_Q$ representations giving rise,
after confinement, to fractionally charged $Q$-hadrons can be
excluded. This is because heavy quarks of this type must be absolutely
stable and, being the same true also for the (lightest) $Q$-hadrons
containing them, it would not be possible to circumvent the exclusion
limits from searches of fractionally charged states~\cite{Perl:2009zz}
by appealing to $Q$-decays.  Absolute stability is obvious in the case
of particles carrying exotic electric charges (e.g.~$\q =1/5, \pi$,
etc.).  They cannot decay into SM particles in force of electric
charge conservation.  Since quarks of this type do not get confined
into hadrons of integer charge, they also cannot get bounded into
neutral hadrons, atoms or molecules.  In other words, their fractional
charge must remain naked.

In case the $Q$'s have less exotic (for example integer) electric
charges, the connection with their absolute stability is less
obvious. However, also in this case it is possible to reach the same
conclusion.  Let us first consider the fundamental particles carrying
color of the SM.  Let us assign $SU(3)_C$ triality $\tau=+1$ to the
fundamental representation $q_j \in \mb[3]$ for the quarks. Then the
reducible representation
$\mb[3]^n = \mb[3]\times \mb[3]\times \mb[3]\dots$, as well as all its
irreducible fragments, have triality $\tau= n$ [mod 3].  For example 
the antisymmetric 2-index representation for the antiquarks
$q_{jk}\in \mb[\bar 3]$ has triality $\tau=2$ since it is an
irreducible fragment of $\mb[3]\times \mb[3] = \mb[\bar 3] + \mb[6]$
(also containing the 2-index symmetric $\mb[6]$), while the
3-index $q_{jkl}\in \mb[3]\times \mb[3] \times \mb[3]$ containing the
totally antisymmetric singlet, the totally symmetric $\mb[10]$, and
two adjoints $\mb[8]$ where the gluons sit, has triality $\tau=3=0$
[mod 3].  SM hadrons are color singlets and are built by contracting
$SU(3)_C$ indices with the totally antisymmetric tensor
$\epsilon_{ijk}$ into invariant index-less tensors.  We can then build
hadrons only from combinations of quarks and gluons of 0-indices [mod
3], e.g.  $q_i q_j q_k$, $q_i q_{jk}$, $q_i q_j q_k q_{l m n}$ etc..
The SM fundamental colored particles have charge $\q(q_j) = -1/3+n$
($n=0,1$) for 1-index (quarks), $\q(q_{ij}) = -2/3 +n$ for 2-index
(antiquarks) and $\q(q_{ijk}) = 0$ for 3-index (gluons).  Therefore, 
any combination of a number of indices multiple of three results in an
integrally charged or neutral state.  So, as is experimentally well
known, all SM hadrons, being color singlets, are integrally charged.
Electric charge conservation then precludes the possibility that
fractionally charged $Q$-hadrons of any type could decay into lighter
SM states.

It is in fact possible to prove a slightly stronger statement: {\sl
  gauge invariant operators inducing decays of exotic heavy quarks $Q$
  are allowed} {\em if and only if} {\sl all color singlet $Q$-hadrons
  are integrally charged or neutral.}  The necessary condition in this
statement is equivalent to the previous result (which can also be
proven in a more direct way and without appealing to electric charge
conservation). As regards the sufficient condition, it is not very
useful since the decay operators which will mandatorily appear could
be so suppressed to render the $Q$'s effectively stable with respect
to all phenomenological consequences. This can happen for example if we
choose $R_Q$'s with particularly large isospin/hypercharge values,
since in this case gauge invariant decay operators would arise at
rather high dimensions.  The proof of the sufficient condition is 
then uninteresting and can be omitted. 

\bibliographystyle{apsrev4-1.bst}
\bibliography{bibliography}

\end{document}